\documentclass[aps,prb,twocolumn,groupedaddress]{revtex4-1}

\usepackage{amsmath}
\usepackage{graphicx}

\usepackage{color}

\begin{document}


\title{Photogalvanic effect in HgTe/CdTe topological insulator due to
  edge-bulk optical transitions}


\author{V.~O.~Kaladzhyan}
\author{P.~P.~Aseev}
\email[]{pavel.p.aseev@gmail.com}
\author{S.~N.~Artemenko} 

\affiliation{Kotel’nikov
	Institute of Radio-engineering and Electronics of Russian Academy of Sciences, Moscow 125009, Russia}
\affiliation{Moscow Institute of Physics and Technology, Dolgoprudny 141700, Moscow region, Russia}


\date{\today}

\begin{abstract}
  We study theoretically 2D HgTe/CdTe quantum well topological insulator (TI) illuminated by circularly polarized light with frequencies
  higher than the difference between the equilibrium Fermi level and the
  bottom of the conduction band (THz range). We show that electron-hole asymmetry results in spin-dependent
  electric dipole transitions between edge and bulk states, and we predict
  an occurrence of a circular photocurrent. If the edge state is tunnel-coupled to a conductor, then the photocurrent can be detected by measuring an electromotive force (EMF)
  in the conductor, which is proportional to the photocurrent.
\end{abstract}


\maketitle

\section{Introduction}
Topological insulators (TIs) became a focus of attention of many condensed matter physicists in recent years, not least due to their possible applications in spintronics and quantum computing. These are materials with time-reversal symmetry and non-trivial topological order, which have an insulating bulk but conducting topologically protected edge/surface states~\cite{KonigJPhysSocJpn2008, HasanKane2010colloqium, QiZhangRevModPhys2011}. Spin-orbit interaction plays a significant role in these materials, and particularly manifests itself in spin-momentum locking of charge carriers in edge/surface states.

Optical excitation is an efficient tool for generating currents in materials. This process has been studied in 3D TIs~\cite{MunizSipePRB2014,MciverNature2012,MisawaPRB2011,JunckPRB2013, HosurPRB2011, JozwiakNatPhys2013, ZhangPRB2010, ParkLouiePRL2012, WuPhysE2012, ZhuPRL2014, PRB-88-045414, PRB-90-205432}. In Ref.~\onlinecite{MunizSipePRB2014} effects in study are due to electric optical transitions between surface states, which are possible in the presence of magnetic field. Besides, in the presence of strong magnetic fields, the electric dipole transitions between Landau levels are also possible~\cite{PRB-88-045414}. In contrast to these papers, here we study 2D TIs and show that the optical generation of the current is possible without magnetic field as well. 

HgTe/CdTe quantum well structures~\cite{BernevigHughesZhangScience2006} are
one of the most well-known 2D TIs. These quantum wells exhibit an inverted band structure if their width exceeds a certain critical value. The inverted band structure and strong spin-orbit interaction give rise to unusual optoelectronic phenomena, e.g. a nonlinear magneto-gyrotropic photogalvanic effect (PGE)~\cite{DiehlPRB2009} (PGE). A circular PGE was also experimentally observed~\cite{WittmannSemiconductor2010} when the sample was illuminated by mid-infrared or terahertz laser radiation. These photocurrents were induced due to direct transitions between different size-quantized subbands or due to indirect (Drude-like) transitions within the lowest size-quantized subband. In both cases the optical transitions responsible for the PGE involve only bulk states. However, in the case of a finite sample size of 2D TI there exist topologically protected helical edge states which form two branches with opposite spins~\cite{BernevigHughesZhangScience2006}. In the recent paper~\cite{arXiv1502.05605} it was predicted that edge states affect bulk magneto-conductivity. The paper focuses on the bulk properties in a strong magnetic field and does not discuss optoelectronic properties of the edge states.  However, it is of interest whether the photocurrent can be induced at the edge states at zero magnetic field.  A PGE due to transitions between edge states of the opposite chiralities has been predicted in Ref.~\onlinecite{ArtemenkoKaladzhyanJETPLett2013}, but electric dipole transitions between them are forbidden by selection rules, and only magnetic dipole transitions are possible in this case. Thus, the direct transitions between the edge states are weak. Unlike Ref.~\onlinecite{ArtemenkoKaladzhyanJETPLett2013}, in this paper we study electric dipole transitions between the edge and bulk states in HgTe/CdTe quantum well 2D TI in zero magnetic field, which lead to the edge currents. To our knowledge, this mechanism of the PGE in TI has not been studied yet. 

Starting from the Bernevig-Hughes-Zhang (BHZ) model  of HgTe/CdTe 2D TI, we find a relation between matrix elements of the edge-bulk transitions. In the general case the electron-hole symmetry is broken, and the probability of transition depends on the spin and, hence, on the chirality. Thus, the transitions will lead to a different population of spin-up and spin-down states and to occurrence of a photoinduced electric current. In order to study this effect, we derive a kinetic equation and then solve it in the quasi-equilibrium approximation. We also propose a way to detect the photoinduced current by coupling 2D TI with a conductor and measuring an electromotive force (EMF) induced in the conductor. 

The paper is organized as follows. In Section~\ref{sec:transitions} we consider optical transitions between edge and bulk states and derive the photoinduced electric current. In section~\ref{sec:emf} we
calculate the EMF that appears in the conductor
tunnel-coupled to the edge state.

Below we set $\hbar=1$, $c=1$.

\begin{figure}
\includegraphics{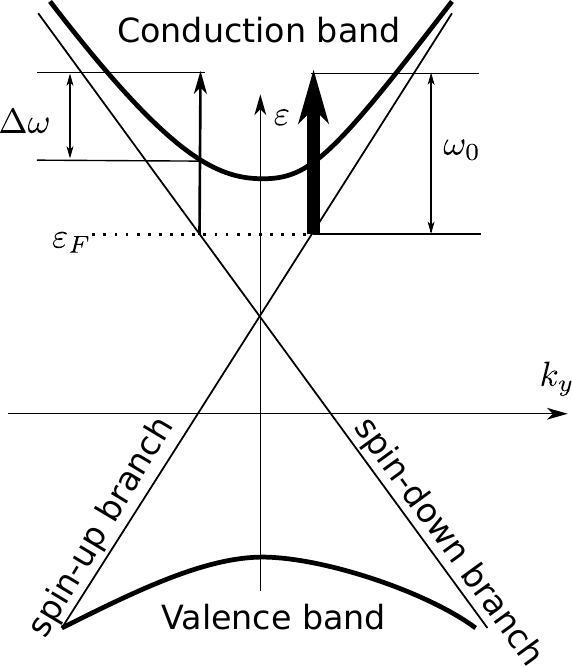}
  \caption{Schematic picture of optical transitions between two edge
    state branches and bulk states. Circularly polarized light of frequency $\omega_0$ exceeding the absorption threshold by $\Delta\omega$ induces transitions between edge and bulk conduction band states. The different thickness of arrows depicts that probability of electric dipole transitions may depend on chirality/spin of the edge electron.}
\label{fig:transitions}
\end{figure}
\section{Optical transitions between edge and bulk states
  \label{sec:transitions}} We consider a HgTe/CdTe quantum well TI
with a conducting helical edge state illuminated by circularly
polarized light with a frequency $\omega_0$ slightly exceeding the
the absorption threshold, so that optical transitions may occur between the
edge state and the bulk conduction band (Fig.~\ref{fig:transitions}).  We assume that the TI is located at $x>0$; $y$-axis is taken along the edge of TI, and $z$-axis
is perpendicular to the 2D TI. 

Both in HgTe and CdTe the relevant bands are s-type band ($\Gamma_6$)
and p-type band split by spin-orbit interaction into a $J=3/2$
($\Gamma_8$) and a $J=1/2$ ($\Gamma_7$) bands. The latter is usually
neglected as it has negligible effects on the band
structure~\cite{BernevigHughesZhangScience2006,NovikPRB2005}. CdTe has
a band-ordering similar to GaAs with $\Gamma_6$ conduction band and
$\Gamma_8$ valence band. In HgTe the usual band-ordering is
inverted. The quantum well subbands derived from the heavy-hole
$\Gamma_8$ are usually denoted by $H_n$, and the subbands derived from
the electron $\Gamma_6$ are denoted by $E_n$.

We describe HgTe/CdTe quantum well TI by the four-band BHZ model~\cite{BernevigHughesZhangScience2006}.

 In the four-component basis consisting of
$\left|E_1\uparrow\right\rangle$, $\left| H_1\uparrow\right\rangle$, $\left|E_1\downarrow\right\rangle$, $\left| H_1\downarrow\right\rangle$ with $m_J=1/2,3/2,-1/2,-3/2$ correspondingly, the Hamiltonian reads:
\begin{multline}
  \label{eqn:BHZ-Ham}
  \hat H_{BHZ} = -\mathcal{D}\hat{\mathbf{k}}^2 +\\ 
  \begin{pmatrix}
 \mathcal{M} - \mathcal{B}\hat{\mathbf{k}}^2&\mathcal{A}\hat{k}_+&0&0\\
  \mathcal{A}\hat{k}_-& -(\mathcal{M} - \mathcal{B}\hat{\mathbf{k}}^2)&0&0\\
  0&0&\mathcal{M} - \mathcal{B}\hat{\mathbf{k}}^2&-\mathcal{A}\hat{k}_-\\
  0&0&-\mathcal{A}\hat{k}_+&-(\mathcal{M} - \mathcal{B}\hat{\mathbf{k}}^2)
  \end{pmatrix},
\end{multline}
where $\hat k_{\pm}= \hat{k}_x \pm i \hat{k_y}$.
Here $\mathcal{A}$, $\mathcal{B}$, $\mathcal{D}$, $\mathcal{M}$ are
material parameters, which depend on quantum well geometry:
$\mathcal{A}>0$, $\mathcal{B}<0$; parameter $\mathcal{M}$ is negative if the
quantum well is in a TI state, and $2|\mathcal{M}|$ is a value of a
band gap in TI. Parameter $\mathcal{D} \neq 0$ if
electron-hole symmetry is broken, and $D=0$ otherwise.

Since the Hamiltonian~(\ref{eqn:BHZ-Ham}) has been obtained in
kp-approximation, and the non-diagonal part of~(\ref{eqn:BHZ-Ham}) linear by quasi-momentum $\hat{\mathbf{k}}$ corresponds to kp-term, the Hamiltonian of a light-matter interaction in
electric dipole approximation reads
\begin{align}
  \hat H_{e-A} &=\;\hat H_1
  \left({{A}}_x - i{{A}}_y \right)+\hat H_1^\dag \left({{A}}_x + i{{A}}_y \right) 
  \label{eqn:light-matter-Ham}\\
    \hat H_1 &=\; |e|\left|H_1\uparrow\right\rangle \left\langle E_1\uparrow\right| -|e| \left|E_1\downarrow\right\rangle \left\langle H_1\downarrow\right|  
\end{align}
where $e$ is the electron charge, $\mathbf{{A}}$ is a vector-potential
of electromagnetic field. In the case of right-hand polarized light (as
defined from the point of view of the source) propagating along
$z$-axis the vector-potential can be represented as
\begin{equation}
  {{A}}_x= \frac{\sqrt{4\pi \mathcal{W}}}{n_{r}\omega_0}\cos\omega_0 t,\quad {{A}}_y = -\frac{\sqrt{4\pi \mathcal{W}}}{n_{r}\omega_0}\sin\omega_0t,\quad {{A}}_z=0,
  \label{eqn:vector-potential}
\end{equation}
where $\mathcal{W}$ is the intensity of light, $n_{r}$ is the refractive index.

Under the illumination by the right-hand circularly polarized light the selection rules allow only those electric dipole transitions from the edge to bulk, which increase angular momentum by $\hbar$. In the Hamiltonian~(\ref{eqn:light-matter-Ham}) these transitions are described by the first term. The conjugate term describes the reverse optical transitions.

In the absence of the boundary, the eigen states of 2D TI are bulk states separated by a gap, and the bulk states at the bottom of the conduction band are formed by $H_1$ states with a zero momentum and a well-defined projection of an angular momentum $m_J=\pm 3/2$.

In a finite-size sample, there appear edge states which are superpositions of $E_1$ and $H_1$ Bloch wavefunctions. These edge states can be found by solving Schroedinger equation with zero
boundary conditions for the Hamiltonian~(\ref{eqn:BHZ-Ham}) in the coordinate representation in $x$-direction and the momentum representation in $y$-direction (see Ref.~\onlinecite{ZhangNaturePhys2009}): 
\begin{multline}
  \psi_{edge,s}\propto \left(is\sqrt{|\mathcal{B}-\mathcal{D}|} \left|E_1, s\right\rangle+\sqrt{|\mathcal{B}+\mathcal{D}|} \left|H_1, s\right\rangle\right)    \times\\ \left(e^{-\lambda_{s,-} x} - e^{-\lambda_{s,+} x} \right),
  \label{eqn:edge-state}
\end{multline}
where $\lambda_{s,\pm}$ are inverse decay lengths for the localized edge states.

In the presence of the boundary, conduction band wavefunctions are distorted near the boundary where they overlap with the edge states. Instead of explicitly calculating the conduction band wavefunctions by solving Schroedinger equation with zero boundary conditions, we will use the time-reversal symmetry and orthogonality conditions for the eigen states of the Hamiltonian. 

The BHZ Hamiltonian and zero boundary condition are invariant under the time-reversal symmetry $\hat \Theta$.
Therefore, if the spin-up eigen state with energy $\epsilon$ is of the form
$\Psi_{bulk,\uparrow}(\epsilon, k_y,x) = 
f_{\epsilon,k_y}(x)\left|E\uparrow\right\rangle + g_{\epsilon,k_y}(x)\left|H\uparrow\right\rangle$, then the spin-down eigen state can be obtained as 
\begin{align*}
\hat{\Theta}\Psi_{bulk,\uparrow}=-
f^*_{\epsilon,-k_y}(x)\left|E_1\downarrow\right\rangle - g^*_{\epsilon,-k_y}(x)\left|H_1\downarrow\right\rangle
  \label{eqn:bulk-state}
\end{align*}
We denote the overlap integrals between the edge and bulk states as:
\begin{align*}
F(\epsilon,k_y) &= \int\limits_0^{+\infty}
\left(e^{-\lambda_-x} - e^{-\lambda_+ x}\right) f_{\epsilon,k_y}(x) dx, \\
G(\epsilon,k_y) &= \int\limits_0^{+\infty} \left(e^{-\lambda_-x} -
  e^{-\lambda_+ x}\right) g_{\epsilon,k_y}(x) dx.
\end{align*}
Mutual orthogonality of edge and bulk states yields the relation between the integrals
\begin{equation}
  \frac{F}{G} = -i \frac{\sqrt{|\mathcal{B}+\mathcal{D}|}}{\sqrt{|\mathcal{B}-\mathcal{D}|}}.
\end{equation}
Thus, both $f$ and $g$ are non-zero, and not only the edge states are superpositions of $E$ and $H$ Bloch wavefunctions with different well-defined projections of total angular momentum, but the bulk conduction states are also their superpositions even at the bottom of the band. Selection rules allow transitions from $\left| E \uparrow\right\rangle$ to $\left| H \uparrow\right\rangle$, and from $\left| H \downarrow\right\rangle$ to $\left|E \downarrow\right\rangle$. Thus, the transitions from the both edge state branches to the conduction band satisfy the selection rules. Calculation of the matrix elements $w_{s}$ of the first term
of the Hamiltonian~(\ref{eqn:light-matter-Ham}) corresponding to these transitions yields the main relation
\begin{equation}
  \frac{|w_{\downarrow}|}{|w_{\uparrow}|} =
  \frac{\sqrt{|\mathcal{B}-\mathcal{D}|}|G|}{\sqrt{|\mathcal{B}+\mathcal{D}|}|F|} = \frac{\mathcal{B}-\mathcal{D}}{\mathcal{B}+\mathcal{D}}.
  \label{eqn:matrix-elements}
\end{equation}
Note, that if electron-hole symmetry is present, the probabilities of optical transitions from the both edge state branches are equal. Electron-hole symmetry implies that the edge states are superpositions of $\left|E,s\right\rangle$ and $\left|H,s\right\rangle$ with equal (up to a phase factor) amplitudes. The same is true for the bulk states (it is shown explicitly in the Appendix~\ref{app:electron-hole-symmetry}). Hence, the transition probability for the both spin-up and spin-down branches will be the same. However, in real samples the electron-hole symmetry is broken i.e. the electron and hole components of the bulk eigen states as well as the edge eigen states are not equal anymore, and the probability of an edge-bulk transition will depend on spin. 

Values of matrix elements for the case of strong electron-hole asymmetry are derived in~Appendix~\ref{app:asymmetry}.

We estimate the ratio of probabilities for typical values of parameters\cite{QiZhangRevModPhys2011}: $\mathcal{A}=365\;\mathrm{meV}\cdot\mathrm{nm}$, $\mathcal{B}=-686\;\mathrm{meV}\cdot\mathrm{nm^2}$,$\mathcal{D}=-512\;\mathrm{meV}\cdot\mathrm{nm^2}$, $\mathcal{M}=-10\;\mathrm{meV}$ corresponding to the quantum well width $d_c=7\;\mathrm{nm}$. In this case $|w_\uparrow|^2/|w_\downarrow|^2\approx 47.4$. Thus, the transitions in case of HgTe/CdTe 2D TI are strongly spin-dependent.

Our consideration can be applicable not only in case of HgTe/CdTe 2D TI but also in case of other 2D TI which can effectively described by BHZ model. One of the interesting examples is a recently predicted all electron TI in InAs double well\cite{ErlingssonPRB2015} which allow easily tune BHZ parameters. In case of this material the band gap $2|\mathcal{M}|$ is of order of $1\;\mathrm{meV}$ and $\mathcal{D}/\mathcal{B}$ is order of $0.5$. These values of parameters correspond to characteristic frequencies $\omega_0$ of order of $100\;\mathrm{GHz}$ and ratio of matrix elements $\frac{|w_{\downarrow}|^2}{|w_{\uparrow}|^2}$ of order of $10$, i.e. the probabilities of transitions is also strongly spin-dependent.

Note, that we used zero boundary conditions (BCs) for 2D TI.  The result~(\ref{eqn:matrix-elements}) does not qualitatively depend on the BCs for the wavefunctions provided they are invariant under time-reversal symmetry and yield helical edge states.  Different BCs are discussed in Ref.~\onlinecite{IsaevPRB2011, MedhiJPhysCondMat2012, MenshovJETPLett2014, EnaldievJETPLett2015}. Although the choice of the boundary conditions does not affect the topological nature and the existence of the edge states, the spectra of bulk and edge states, and the eigenstates themselves depend on the BCs. Particularly, the matrix elements of the transitions may depend on the BCs. However, in our approach only the Eq.~(\ref{eqn:edge-state}) depends on the BCs. In general case the amplitudes of $E_1$ and $H_1$ Bloch functions will be different, but if the electron-hole symmetry is broken these amplitudes will remain still unequal. Further, in order to obtain our main result~(\ref{eqn:matrix-elements}) we exploit their inequality, time-reversal symmetry and mutual orthogonality of eigenstates. Thus, we believe that the result does not qualitatively depend on the BCs provided they are invariant under time-reversal symmetry and yield helical edge states.

If the light is incident in an arbitrary direction
$\mathbf{n}^{\theta,\phi}=(\cos\phi\sin \theta, \sin \phi\sin\theta,
\cos \theta)$, then the matrix elements $w_s^{\theta,\phi}$ can be
obtained by replacing vector potential in the light-matter interaction Hamiltonian~(\ref{eqn:light-matter-Ham}) with its projection on the TI plane(for details see~Appendix~\ref{app:angle}):  
\begin{equation}
  w_s^{\theta,\phi} = \frac{w_se^{-i\phi}(1+\cos\theta) + w_{-s}^*e^{i\phi}(1-\cos\theta)}{2}.
\label{eqn:angle}
\end{equation}
Below we will use an expression for the value of the squared matrix element averaged over the direction:
\begin{equation}
  \left\langle w^2_s \right\rangle_{\theta,\phi} = \frac{1}{3}\left(|w_s|^2 + |w_{-s}|^2 \right)
  \label{eqn:averaged-w}
\end{equation}

Kinetic equations for distribution functions of electrons in the edge
state $n(\varepsilon)$ and in the conduction band
$N(\varepsilon)$ can be written as
\begin{multline}
  \frac{dn_s(\varepsilon)}{dt} =
  -\frac{n_s(\varepsilon)-N_s(\varepsilon+\omega_0)}{\tau_{ind,s}(\varepsilon,\varepsilon+\omega_0)}\mathcal{W}
  +\\\int
  \frac{N_s(\varepsilon+\omega)\left[1-n_s(\varepsilon)\right]}{\tau_{sp}(\varepsilon+\omega,\varepsilon)}d\omega 
  \label{eqn:kinetic-equation} +\\\frac{n_{-s}(\varepsilon)-n_s(\varepsilon)}{\tau_e}
\end{multline}
where $\mathcal{W}\tau_{ind}^{-1}$ is the rate of transitions induced by
illumination, $\tau^{-1}_{sp}(\varepsilon',\varepsilon)$ is the rate of
spontaneous transitions between the conduction bulk state with energy
$\varepsilon'$ and the edge state with energy $\varepsilon$,
i.e. recombination rate; $\tau_e$ is the spin relaxation time for the edge electrons. Since there is still discussion in the literature\cite{GusevPRB2014, MaciejkoPRL2009, TanakaPRL2011, LundePRB2012, VayrunenPRB2014} which mechanism agrees better with experimental data, in this paper we introduce this time assuming that in any realistic system it is finite. 

Here we write these  kinetic equations~(\ref{eqn:kinetic-equation}) phenomenologically, and more
rigorous derivation based on the Keldysh technique is given in the Appendix~\ref{sec:ke-appendix}. The rate of the transitions
induced by illumination can be related to the matrix elements using
the Fermi golden rule
$$
\tau_{ind,s}^{-1} =
8\pi^2\tilde{\nu}_{C,k_y}(\varepsilon+\omega_0)\frac{|w_s|^2}{n_{r}^2\omega_0^2},
$$
where $\tilde{\nu}_{C,k_y}$ is the density of states in the conduction band with a fixed $k_y$: $$\tilde{\nu}_{C,k_y} = \sum\limits_{k_x} \delta\left(\varepsilon -\epsilon_{k_x,k_y} \right) = \frac{\sqrt{m}L_x}{\sqrt{2}\pi\sqrt{\varepsilon-|\mathcal{M}|-k_y^2/(2m)}},$$
where $m$ is the effective mass of conduction band electrons and $L_x$ is the length of 2D TI in the $x$ direction. Note that summation over $k_x$ in the definition of $\tilde{\nu}_{C, k_y}$ instead of summation over both $k_x$ and $k_y$ reflect the fact, that the transitions are vertical, i.e. $k_y$ projection of momentum conserves. 

In order to deduce a relation between induced and spontaneous transition rates one can use a detailed balancing condition similar to that for Einstein
coefficients for discrete levels\cite{amnon1989quantum}. The factors $\tau_{sp,ind}$ in the kinetic equation do not depend on the illumination and environment, since they are intrinsic properties of the 2D TI. Therefore, the kinetic equation~(\ref{eqn:kinetic-equation}) should remain valid if we put the system in thermal equilibrium with black body radiation. In this case the distribution functions of the edge and bulk electrons are the equilibrium Fermi function with the same Fermi level, and photons have the Bose distribution. The detailed balancing between the states with energy
$\varepsilon$ and $\varepsilon'=\varepsilon+\omega$ for an arbitrary
$\omega$ yields
\begin{multline}
  \left[n_0(\varepsilon) - n_0(\varepsilon+\omega)\right]\left\langle\tau_{ind,s}\right\rangle_{\theta,\phi} \frac{d\mathcal{W}_{eq,+}}{d\omega}+\\\left[n_0(\varepsilon) - n_0(\varepsilon+\omega)\right]\left\langle\tau_{ind,-s}\right\rangle_{\theta,\phi} \frac{d\mathcal{W}_{eq,-}}{d\omega}=\\
  \frac{n_{0}(\varepsilon+\omega)\left[1-n_{0}(\varepsilon)\right]}{\tau_{sp}},
  \label{eqn:detailed-balancing}
\end{multline}
where $\langle \tau_{ind,s}^{-1}\rangle_{\theta,\phi}$ is the induced
transition rate averaged over the direction of an incident equilibrium
photon, $n_0=\frac{1}{2}\left(1-\tanh \frac{\varepsilon}{2T}\right)$
is the equilibrium Fermi distribution, and
$\frac{d\mathcal{W}_{eq,+(-)}}{d\omega} =
\frac{n_{r}^3\omega^3}{2\pi^2}N_{ph}(\omega)$ is the spectral density of equilibrium right(left) polarized illumination with
the photon distribution function
$N_{ph}(\omega)=\frac{1}{2}\left(\coth\frac{\omega}{2T}-1\right)$. The
value of the induced transition rate $\langle\tau_{ind,s}^{-1}\rangle$ averaged over direction can be calculated
using~(\ref{eqn:averaged-w})
$$
\left\langle \tau_{ind}^{-1} \right\rangle_{\theta,\phi} =
\frac{\tau_{ind,s}^{-1}+\tau_{ind,-s}^{-1}}{3}
$$
Finally, we obtain the expression for the spontaneous illumination rate
$\tau_{sp}^{-1}$ from~(\ref{eqn:detailed-balancing})
\begin{equation}
  \tau_{sp}^{-1} = 
  \frac{8}{3}\left(|w_s|^2+|w_{-s}|^2 \right)n_{r}\omega\tilde{\nu}_{C,k_y}(\varepsilon+\omega)
\end{equation}

We solve kinetic equation~(\ref{eqn:kinetic-equation}) in the
quasi-equilibrium approximation, i.e. assuming that the distribution
function of edge electrons with spin $s$ is a Fermi distribution with
a quasi Fermi levels $\varepsilon_F+\mu_{s}$, and, similarly, the
distribution function of conduction bulk electrons with spin $s$ is a
Fermi distribution with a quasi Fermi level $|\mathcal{M}|+\zeta_s$
($|\mathcal{M}|$ is the bottom of the conduction band). The quasi-equilibrium approximation can be justified if the
energy relaxation times in the edge and bulk states are much shorter than the life-time of the excess photogenerated electrons. Since the results will depend on whether the initial Fermi level is above or below the Dirac point we consider both these cases. 
\subsection{Fermi level above Dirac point: absorption without photocurrent}
\begin{figure*}
	\begin{tabular}{ccc}
		\includegraphics{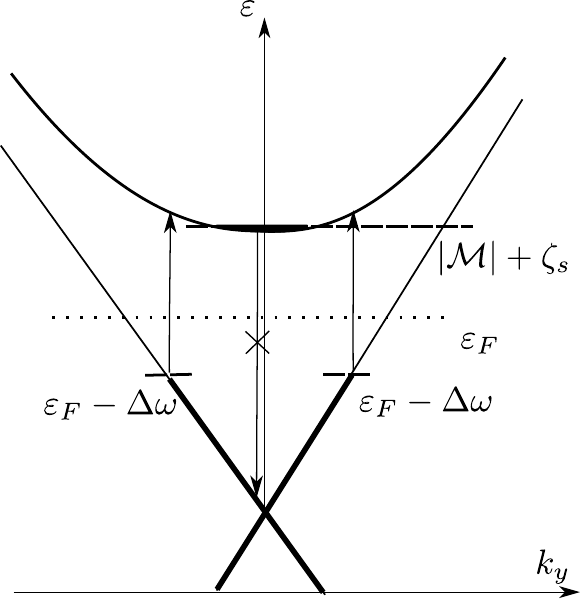}&
		 \includegraphics{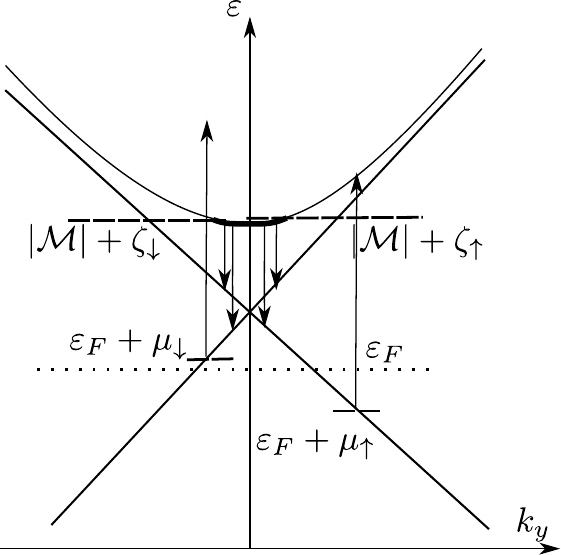}& 
		 \includegraphics{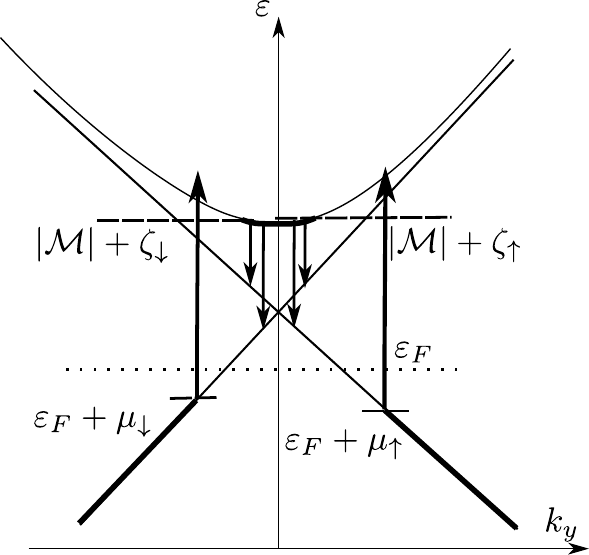} \\
		(a)&(b)&(c)
	\end{tabular}
	\caption{(a) Schematic picture of quasi-Fermi levels if Fermi level is above the Dirac point. All the spin up and spin down edge electrons with energies from $\varepsilon_F - \Delta\omega$ to $\varepsilon_F$ are moved to the conduction band. States near the Dirac point are occupied, and, hence, spontaneous transitions from bulk to edge states are not allowed. The electric current of the edge electrons $j=G_0(\mu_\uparrow-\mu_\downarrow)=0$;
		(b) Schematic picture of quasi-Fermi levels if Fermi level is below the Dirac point and for low intensities $\mathcal{W}<\mathcal{W}_\downarrow-\mathcal{W}_\uparrow$ (region I on the~Fig.~\ref{fig:plot}). All the spin-up electrons with energies from $\varepsilon_F-\Delta\omega$ to $\varepsilon_F$ are excited by the light, and spin-down electrons in the conduction band appear mainly due to spin relaxation. Recombination of bulk spin-down electrons shifts the quasi-Fermi level of spin-down electrons above the initial Fermi level.
		(c)  Schematic picture of quasi-Fermi levels for high intensities $\mathcal{W}>\mathcal{W}_\downarrow/2$ (region II on the~Fig~\ref{fig:plot}). Almost all spin-up and spin-down electrons with energies from $\varepsilon-\Delta\omega$ to $\varepsilon$ are excited by the light($\mu_\uparrow\approx \mu_\downarrow\approx-\Delta\omega$). The spin imbalance at the edge states decays with intensity of the light as $\delta \mu \propto \mathcal{W}^{-2}$
		}
		
	\label{fig:quasiresults}
\end{figure*}

The electrons in the conduction band in the quasi-equilibrium approximation lie in the bottom of the conduction band and they can recombinate only with empty states in the vicinity of the Dirac point. If the Fermi level is above the Dirac point (see~Fig.~\ref{fig:quasiresults}a) then all the states near Dirac point are occupied $1-n_s(\varepsilon) = 0$, and spontaneous transition term in~(\ref{eqn:kinetic-equation}) vanishes.

After integrating the kinetic equation~(\ref{eqn:kinetic-equation})
over energies we obtain a relation
between quasi-Fermi levels
$$
8\pi\frac{|w_s|^2 L_x\sqrt{2m}}{n_r^2\omega_0^2 } \sqrt{\mu_s + \Delta\omega}\mathcal{W} = \frac{\mu_{-s}-\mu_s}{\tau_e},
$$
where $\Delta \omega$ is the difference between the light frequency and the absorption threshold $\Delta\omega = \varepsilon_F+\omega_0 - |\mathcal{M}| - k_y^2/(2m)$ (see~Fig.~\ref{fig:transitions})
The only solution is $\mu_{-s} = \mu_s = -\Delta \omega$. Almost all the electrons with energies from $\varepsilon_F-\Delta \omega$ to $\varepsilon_F$ are moved to the conduction band by illumination($\mu_\uparrow\approx -\Delta\omega$). They equilibrate in the conduction band and remain there, since the edge states near the Dirac point are occupied. The electric current $j=G_0 (\mu_\uparrow - \mu_\downarrow) = 0$, where $G_0=e^2/h$ is the conductance quantum.

\subsection{Fermi level below Dirac point: non-zero photocurrent}

The situation differs if the Fermi level is below the Dirac point. In this case the edge states in the vicinity of the Dirac point are not occupied and the spontaneous transitions from the conduction band to edge states are allowed.
Integration of~(\ref{eqn:kinetic-equation}) yields
\begin{align}
  &\mathcal{W}\sqrt{\mu_s + \Delta\omega}
  = \frac{\nu_C}{\nu_e}\frac{\mathcal{W}_s}{\sqrt{\Delta \omega}}\left(\zeta_s + \tau_0\frac{\mu_{-s}-\mu_s}{\tau_e} \right) ,
  \label{eqn:quasi-fermi}\\
  &\mathcal{W}_{s} = \frac{\nu_e}{3\pi\nu_C}\omega_0^2\left(|\mathcal{M}|-\varepsilon_D\right)\sqrt{\frac{m v_{TI}^2}{2}\Delta\omega}\frac{|w_s|^2+|w_{-s}|^2}{|w_s|^2}n_{r}^{3} \notag,\\
  &\tau_0^{-1} = \frac{8L_x  n_rmv_{TI} \left(|\mathcal{M}|-\varepsilon_D \right)\left(|w|_s^2+|w|_{-s}^2 \right)}{3} ,\notag
\end{align}
where $\Delta \omega = \varepsilon_F + \omega_0 -
|\mathcal{M}|$ (see Fig.~\ref{fig:transitions}). Here $\nu_e$ and $\nu_C$ are the densities of the edge states and the conduction bulk states correspondingly. Note that in contrast to $\tilde{\nu}_{C,k_y}$ defined above,  $\nu_C=\sum\limits_{k_y} \tilde{\nu}_{C,k_y} $ is the 2D density of states in which summation over both components of momentum is performed.
 Another relation results from the conservation law for the
number of particles
\begin{align}
  \nu_{e}(\mu_\uparrow + \mu_\downarrow) +\nu_C (\zeta_\uparrow + \zeta_\downarrow) =0.
  \label{eqn:number-conservation}
\end{align}
In the stationary regime neither spin nor charge accumulates, and the same number of transitions per unit time occurs from spin-up states to spin-down states and vice versa. Therefore we can equate spin-relaxation rates of the edge and bulk electrons and obtain another quasi-equilibrium condition:
\begin{align}
\nu_{e}\frac{\mu_s - \mu_{-s}}{\tau_e} =  \nu_{C}\frac{\zeta_{-s} - \zeta_{s}}{\tau_C},
\label{eqn:spin-relaxation}
\end{align}
where $\tau_C \ll \tau_e$ is the spin relaxation time for the conduction band electrons. 
\begin{figure}
	\includegraphics[width=0.9\columnwidth]{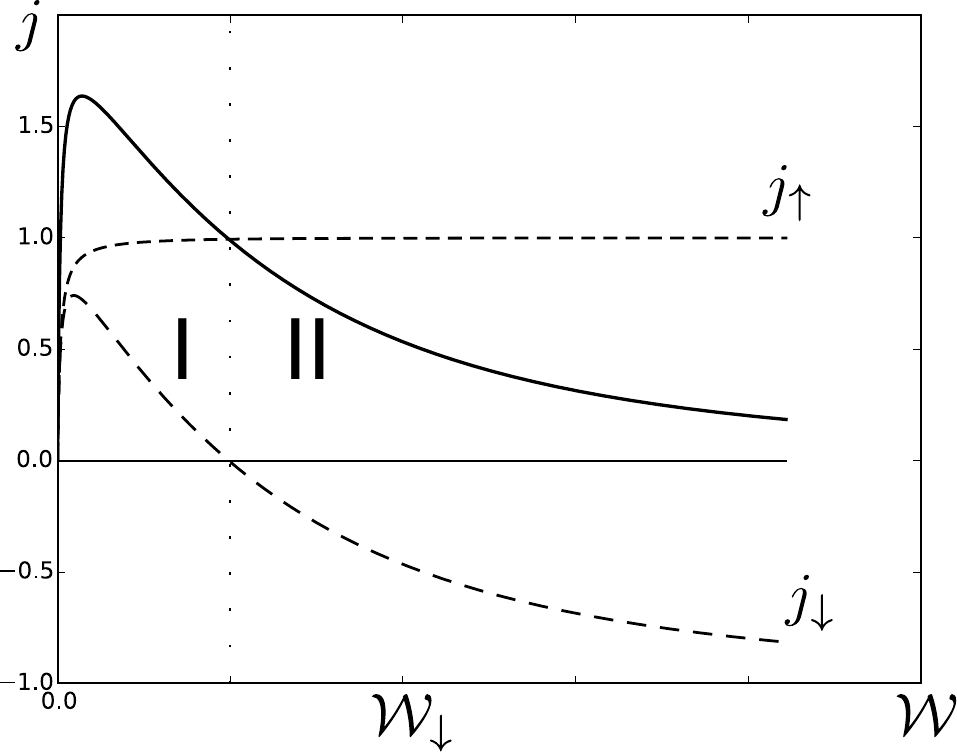}
	\caption{ Dependence of photocurrent (solid line) and currents of spin-up/spin-down electrons (dashed lines) on light intensity if the initial Fermi level is below the Dirac point. The currents of each branch are proportional to the corresponding quasi-Fermi level shift. For the region I ($\mathcal{W}<\mathcal{W}_\downarrow/2$) edge branches contribute to net current with the same sign, while in the region II ($\mathcal{W}>\mathcal{W}_\downarrow/2$) the current of spin-down electrons changes its sign. Currents are measured in the units of $G_0 \Delta\omega$.}	
	\label{fig:plot}
\end{figure}

An important limiting case is when one can neglect spin relaxation of the edge electrons assuming that $\tau_e \gg \tau_C$ and both times are great enough. In this case equation~\ref{eqn:number-conservation}--\ref{eqn:spin-relaxation} yield $\zeta_s=\zeta_{-s}=-\frac{\nu_e}{\nu_C}\frac{\mu_\uparrow+\mu_\downarrow}{2}$, and the equation~(\ref{eqn:quasi-fermi}) can be solved analytically. 
\begin{align}
& \mu_0 = \frac{\mu_\uparrow+\mu_\downarrow}{2} = \frac{\mathcal{W}^2 - \sqrt{\mathcal{W}^4 +2\mathcal{W}^2\left(\mathcal{W}_\uparrow^2+\mathcal{W}^2_\downarrow\right)} }{(\mathcal{W}_\uparrow^2+\mathcal{W}_\downarrow^2)}\Delta\omega\\
  & \delta \mu = \frac{\mathcal{W}_\uparrow^2-\mathcal{W}^2_\downarrow}{\Delta \omega \mathcal{W}^2}\mu_0^2
\label{eqn:delta-mu}
\end{align}
An electric current $j=G_0 \delta \mu$ arises in the edge state, where $G_0 = e^2/h$ is the conductance quantum. If the intensity of light is small $\mathcal{W} \ll \mathcal{W}_\uparrow, \mathcal{W}_\downarrow$ (Fig~\ref{fig:quasiresults}b), mainly the spin-up electrons are excited by the light. The spin relaxation time for the bulk states is much shorter than the life-time of photogenerated electrons, hence, spin of the photogenerated electrons relaxes, and there appear additional spin-down electrons in the conduction band, whose recombination rate turns out to be greater than the excitation rate of spin-down edge electrons. Thus, the quasi Fermi level for spin-up electrons is below $\varepsilon_F$, and the quasi Fermi level for spin-down electrons is above it. The currents of electrons with the opposite spin contribute to the total electric current with the same sign (see Fig~\ref{fig:plot}a,~Fig~\ref{fig:plot}c), and the total current $j\approx 2G_0 \frac{W_\uparrow^2-W_\downarrow^2}{W_\uparrow^2+W_\downarrow^2}\Delta\omega$.

If the intensity of light $\mathcal{W} \gg \mathcal{W}_\uparrow, \mathcal{W}_\downarrow$ (Fig.~\ref{fig:plot}b), all the edge electrons with energies from $|\mathcal{M}|-\omega_0$ to $\varepsilon_F$ are excited to the conduction band, and the quasi Fermi levels of the edge electrons saturate at the value $\mu_s=-\Delta \omega$. The currents $j_s = G_0 \mu_s$ are almost equal, but contribute to the total current with opposite signs. Thus, the total current decreases with the increase of the intensity as $j \propto \mathcal{W}^{-2}$. For an HgTe/CdTe quantum well with width $d_c=7\;\mathrm{nm}$, and the sample size of order $L\sim 1\;\mathrm{\mu m}$, the quasi Fermi level for spin-up electrons saturates at $\mathcal{W}\sim\mathcal{W}_\uparrow \sim 10^{-9}\;\mathrm{W/cm^2}$. The quasi Fermi level for spin-down electrons saturates at $\mathcal{W}_\downarrow \sim 10^{-8}\;\mathrm{W/cm^2}$

However, in case of extremely weak intensities spin relaxation rate turns out to be comparable with transitions rates and hence cannot be neglected. We can solve~(\ref{eqn:quasi-fermi})--(\ref{eqn:spin-relaxation}) analytically assuming $\mathcal{W}<\mathcal{W}_\uparrow\ll\mathcal{W}_\downarrow$ and obtain 
$$
j = G_0 \Delta\omega \frac{\mathcal{W}^2}{4\mathcal{W}_\uparrow^2}\frac{\tau_e^2}{\tau_C^2}\left(
1- \sqrt{1+16\frac{\mathcal{W}_\uparrow^2}{\mathcal{W}^2}\frac{\tau_c^2}{\tau_e^2}}
\right)
$$

The full curve for the dependence of photocurrent on intensity is sketched on the Fig.~\ref{fig:plot}.

\begin{figure}
	\includegraphics{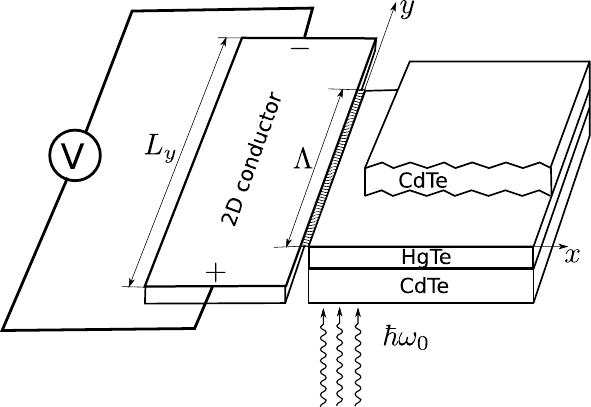}
	\caption{A proposal to detect photocurrent in the edge state: 2D conductor of length $L_y$ is coupled to the edge state via the tunnel contact of length $\Lambda$. The EMF appears between the opposite ends of the 2D conductor. }
	\label{fig:system}
\end{figure}
\section{Tunnel contact to an external circuit \label{sec:emf}}
In the previous section we showed that circularly polarized
illumination induces an electric current at the edge of HgTe/CdTe
quantum well TI. In order to observe the effect one should
connect the sample to an external circuit. We consider a system of
HgTe/CdTe tunnel-coupled by a contact of length $\Lambda$ to a 2D metal
conductor~(see~Fig.~\ref{fig:system}).

The Hamiltonian of the system reads
\begin{align}
  \hat H = \hat H_{TI} + \hat H_{2D} + \hat H_{tun} \label{eqn:total-Hamiltonian}
\end{align}
In order to describe the edge state we use an effective edge state
Hamiltonian corresponding to the linear edge state
spectrum
\begin{align}
  \hat H_{TI} = \sum\limits_s \int \hat   \psi_{s}^\dag(y) (-iv_{TI}s\partial_y+\varepsilon_{D}) \hat \psi_s(y) dy,
\end{align}
where $\hat \psi$ is the effective field operator for an electron in the edge
states, $\varepsilon_{D}$ is the energy of Dirac point measured from the middle of the band gap, $v_{TI}$ is the velocity of edge electrons. We assume that illumination results in quasi-equilibrium
occupation numbers of electrons corresponding to the quasi Fermi levels, and the tunnel
coupling is weak that it does not affect the optical transitions, so
the results of the previous section are applicable. The Hamiltonian of
the 2D conductor reads
\begin{align}
  \hat H_{2D} = \hat{H}_{2D,0} + \sum\limits_s\int d^2 \mathbf{r}\; \hat\Psi_{s}^\dag V_{imp}(\mathbf{r}) \hat \Psi_{s},
\end{align}
where $\hat H_{2D,0}$ is the Hamiltonian of free 2D electrons, $\hat\Psi$ is the field operator in the
2D conductor.  Here we take into account a random
delta-correlated potential of impurities $V_{imp}$ characterized by a mean scattering time
$\tau_{imp}$.

The tunneling Hamiltonian reads:
\begin{equation}
  \hat H_{tun} = \sum\limits_s\int dy\; \hat \psi_s^\dag(y) \mathcal{T} \hat\Psi_s(x=0,y) + H.c.\label{eqn:tunneling-Hamiltonian}
\end{equation}
where $\mathcal{T}$ is a matrix element of tunneling, and we
assume that the tunneling is momentum-conserving.

We start from the
Hamiltonian~(\ref{eqn:total-Hamiltonian})--(\ref{eqn:tunneling-Hamiltonian}),
and then derive equations for Keldysh Green functions $G^{R(A),K}$
taking into account the tunneling Hamiltonian and the impurity potential as perturbations. The self energy of 2D electrons in the
conductor resulting form the tunneling reads
\begin{align*}
  \Sigma^{R(A),K}_{tun}(\mathbf{r}_1,\mathbf{r}_2) = |\mathcal{T}|^2\delta(x_1)\delta(x_2)G_{TI}^{R(A),K}(y_1-y_2)\label{eqn:self-e-tunneling}
\end{align*}
The Green functions of the electrons in the edge state electrons can be obtained by
solving the corresponding Dyson equations:
\begin{align}
  G_{TI,s}^{R(A)}(\varepsilon, p_y) &=\; \left(\varepsilon - \varepsilon_{p_y,s}^{} + i\Gamma\right)^{-1}\\
  G_{TI,s}^K(\varepsilon,p_y) &=\; -2\pi i \delta_\Gamma(\varepsilon-\varepsilon^{}_{p_y,s})\tanh\frac{\varepsilon-\varepsilon_F^{(TI)}-\mu_s}{2T}
\end{align}
where
$\Gamma$ is the inverse life-time of electrons in the edge state, which consists
of contributions from tunneling and optical transitions. The contribution from the optical transitions is determined by self energy operator $i\Gamma = \sum_{k_y} (\Sigma^R-\Sigma^A)$~(see~(\ref{eqn:Gamma-ind}),~(\ref{eqn:Gamma-sp})). Besides, some mechanisms of spin-relaxations, e.g. coupling to multiple puddles\cite{VayrunenPRB2014} also contribute to $\Gamma$, so here we introduce it phenomenologically. Note, that the exact value of $\Gamma$ does not affect the final results. The Lorenz-type factor
$\delta_\Gamma(x)=\frac{1}{\pi}\frac{\Gamma}{x^2+ \Gamma^2}$ describes
broadening of electronic states due to their finite life-time.

The derivation of kinetic equation for
quasi-classical distribution function $f_s(\mathbf{r},\mathbf{p})$ in
the 2D conductor is straightforward~\cite{Keldysh1965diagram} and gives
\begin{multline}
  \left[\partial_t + (\mathbf{v},\nabla) \right]f_s +\\ 2\pi|\mathcal{T}|^2 \delta(x)\theta\left(\frac{\Lambda}{2}-|y| \right) \delta_\Gamma\left(\varepsilon^{}_{p_y,s} - \varepsilon^{(2D)}_{\mathbf{p},s} \right)\left[f_s-n_s\right] =\\ \frac{\bar{f}_s-f_s}{\tau}
\end{multline}
Here $n_{s}$ is a distribution function for the edge state electrons the TI, and the Heaviside
step-function $\theta$ restricts the length of the contact to
$\Lambda$.

If the TI is illuminated by circular-polarized light, the quasi Fermi level of spin-up electrons is lower than the quasi Fermi level of spin-down electrons. In a stationary regime the current through the tunnel contact should be equal to zero, therefore, the Fermi level in the 2D conductor is exactly in the middle between the quasi Fermi levels in the edge states of TI. However, the zero tunnel current consists of a spin-up current from the TI to the conductor and a spin-down current from the conductor to TI. Electrons of the opposite spins in the TI are of the opposite chiralities and the tunneling is assumed to conserve momentum,so it should result in an electron drift and appearance of the counterbalancing EMF in the conductor. We consider a stationary case in which tunneling leads to charge redistribution, and, hence, the extra charge induces an electric field.The electric field $\mathbf{E}$ is related with the extra charge $-|e|\delta \rho$ by the Poisson equation:
\begin{align}
  \mathrm{div}\; \mathbf{E} = -4\pi|e|\delta \rho\label{eqn:Poisson-equation},
\end{align}
where $\delta \rho$ is the deviation of electron density from its equilibrium value, and the full electron density can be expressed in terms of the distribution
function as
\begin{align}
  \rho = \sum\limits_{s} \int \frac{d^2 \mathbf{p}}{(2\pi)^2} \bar{f}_s
\end{align}

The kinetic equation that takes into account the electric field in the
conductor reads
\begin{align}
  & \left[\partial_t + (\mathbf{v},\nabla) \right]f_s + \Pi_s + \frac{f_s-\bar{f}_s}{\tau} -|e|\mathbf{v}\mathbf{E}\frac{\partial f_s}{\partial \varepsilon}  = 0,
      \label{eqn:kinetic-lead}\\
  &  \Pi_s = \theta\left(\frac{\Lambda}{2}-|y|
    \right) 2\pi|\mathcal{T}|^2  \delta(x) \delta_\Gamma\left(\varepsilon^{}_{p_y,s} -
    \varepsilon^{(2D)}_{\mathbf{p},s}
    \right)\left[f_s-n_s\right] 
    \label{eqn:tunneling-source}
\end{align}
Here $\Pi_s$ denotes the term responsible for the tunneling.  Since
the electric field $\mathbf{E}$ arises due to the deviation of the
distribution function $f$ from its equilibrium unperturbed
value $f_0$, we can replace $f_s$ in the field term with $f_0$,
neglecting second-order corrections.

We consider a model where the 2D conductor is a narrow strip spanning from $x=-L_x/2$
and $x=L_x/2$ and is infinite in $y$-direction, and the contact
is in the middle of the strip at $x=0$. For simplicity we
assume that the width $L_x$ of the conductor is smaller than the scattering
length $v_F \tau$. In a more general case the conductor can be qualitatively considered as
the same narrow strip with a bypass resistance. Under these
assumptions one can take into account only the electric field along the contact
$E_y \gg E_x$.

We solve the kinetic equation together with the Poisson equation assuming quasi-neutrality and treating the
angle-averaged distribution function $\bar{f}_s$ in the impurity term
self-consistently (for details see~Appendix~\ref{sec:emf-appendix}).

Finally, we obtain the EMF as the integral over the conductor length of the electric field averaged in
the transversal direction
$\langle E_y \rangle_x = L_x^{-1}\int\limits_{-L_x/2}^{L_x/2} E_y
\;dx$:
\begin{multline}
  |e|\mathcal{E} = \int\limits_{-L_y/2}^{L_y/2} |e|\langle E_y \rangle_x \; dy =\\ -\frac{|\mathcal{T}|^2}{v_F}\frac{2(\varepsilon_F^{(TI)}-\varepsilon_{D})}{\sqrt{\left(p_Fv_{TI}\right)^2-(\varepsilon_F^{(TI)} - \varepsilon_{D})^2}} \frac{\Lambda}{L_x}\frac{\delta\mu}{v_{TI}p_F},
\end{multline}
where $p_F$, $v_F$ are the Fermi momentum and velocity in the conductor, and $\delta \mu$ is the
difference between quasi Fermi levels given by~(\ref{eqn:delta-mu})

\section{Conclusions}
In summary, we considered electric dipole optical transitions from
helical edge states of HgTe/CdTe TI illuminated by circularly
polarized light to the bulk conduction band. If electron-hole symmetry is broken which typically is the case, these optical transitions are strongly spin-dependent. 

 This gives rise to a circular electric current in the edge state if the Fermi level is below the Dirac point. The value of the photocurrent reaches maximum and then decreases with the growth of the light intensity. It is worth noting that although the overlap between wavefunctions of the edge and bulk states determines the time required for stationary regime to settle in, the magnitude of the current in this regime  does not
depend on the overlap, and,
hence, we anticipate that the effect can be observed even in samples where this overlap
is small.  We showed that the photocurrent can be detected electrically
by measuring EMF in the conductor coupled to the edge state of the TI.

\section*{Acknowledgments}
We are grateful to K.~E.~Nagaev for helpful comments and discussions. 
This work was supported by Russian Foundation for Basic Research, and by the program of Russian Academy of Sciences.

\appendix
\section{Optical transitions in the case of electron-hole symmetry \label{app:electron-hole-symmetry}}

In this section the matrix elements of the optical transitions are explicitly calculated for the case of electron-hole symmetry $\mathcal{D}=0$. In this case, the Hamiltonian~(\ref{eqn:BHZ-Ham}) with zero boundary conditions yields the edge state wavefunction
\begin{align}
&\psi_{edge,s}(x,k_y) = \Psi_{edge,s}
\left(e^{-\lambda_{s,-} x} - e^{-\lambda_{s,+} x} \right), \notag\\
&\Psi_{edge,\uparrow} \propto \begin{pmatrix}
i\\
1\\0\\0
\end{pmatrix}, \quad \Psi_{edge,\downarrow}\propto
\begin{pmatrix}
0\\0\\
-i\\
1\\
\end{pmatrix},
\label{eqn:sym-edge-state}
\end{align}
where $\lambda_+ \approx \frac{\mathcal{A}}{|\mathcal{B}|}$, $\lambda_- \approx \frac{|\mathcal{M}|}{\mathcal{A}} \ll \lambda_+$. Here we assumed that $\mathcal{A}^2\gg \mathcal{B}\mathcal{M}$, which is the case for the typical parameters of TI. Solving Schroedinger equation with zero boundary condition for $\epsilon-|M|\ll |M|$, $k_y=0$ we obtain
bulk wavefunctions near the bottom of the conduction band: 
\begin{align}
\psi_{bulk,\uparrow}(\epsilon) \propto 
\begin{pmatrix}
\cos k_x x + \frac{k_x\mathcal{A}}{2\mathcal{M}}\sin k_x x\\
i\cos k_x x -\frac{2i\mathcal{M}}{k_x\mathcal{A}}\sin k_x x\\
0\\
0
\end{pmatrix}   - 	\begin{pmatrix}
1\\
i\\
0\\
0
\end{pmatrix}e^{-\kappa x},\label{eqn:bulk-up}\\
\psi_{bulk,\downarrow}(\epsilon) \propto
\begin{pmatrix}
0\\
0\\
\cos k_x x + \frac{k_x\mathcal{A}}{2\mathcal{M}}\sin k_x x\\
-i\cos k_x x +\frac{2i\mathcal{M}}{k_x\mathcal{A}}\sin k_x x
\end{pmatrix}   - 	\begin{pmatrix}
0\\
0\\
1\\
-i
\end{pmatrix}e^{-\kappa x}\label{eqn:bulk-down},
\end{align}
where $k_x \approx \frac{\sqrt{\epsilon^2-\mathcal{M}^2}}{\mathcal{A}}\ll \lambda_-$ and $\kappa\approx\frac{\mathcal{A}}{|\mathcal{B}|}\approx \lambda_+$. The first terms in~(\ref{eqn:bulk-up})--(\ref{eqn:bulk-down}) correspond to the superposition of incident and reflected waves, while the second terms correspond to the parts of wavefunctions localized near the boundary. In the presence of the boundary the angular momentum is ill-defined. Although the wavefunctions of the conduction band behave like $|H_1, m_j =3/2\rangle$ and $|H_1, m_j=-3/2 \rangle$ with a well-defined angular momentum far away from the boundary at $x\gg k_x^{-1}$, the overlap integral between edge and bulk states is dominated by small distances from the boundary $x\sim \lambda_{-}^{-1}$, where these wavefunctions are superpositions of $|E_1\rangle$, $|H_1\rangle$ with equal (up to a phase factor) amplitudes:
\begin{align}
\psi_{bulk,\uparrow}(\epsilon) \sim 
\begin{pmatrix}
1\\
i\\
0\\
0
\end{pmatrix}\left(1-e^{-\kappa x}\right),\\
\psi_{bulk,\downarrow}(\epsilon) \sim
\begin{pmatrix}
0\\
0\\
1\\
-i
\end{pmatrix}\left(1-e^{-\kappa x}\right),
\end{align}
Thus, the matrix elements are equal up to a phase factor $|w_{\uparrow}|^2 = |w_{\downarrow}|^2 \sim \frac{1}{L_x\lambda_-^3}$.

\section{Optical transitions in the case of strong electron-hole asymmetry\label{app:asymmetry}}
In this section we analytically derive matrix elements of the optical transitions in
the case of strong electron-hole asymmetry, i.e. $|\mathcal{B}-\mathcal{D}|\ll\mathcal{B}$.
For the simplicity we consider transitions between the Dirac point $k_y=0$ and the bottom of the conduction band $\varepsilon-|\mathcal{M}| \ll |\mathcal{M}|\sqrt{\mathcal{B}^2-\mathcal{D}^2}/B $. We also assume that typically $\mathcal{A}^2\gg\mathcal{B}\mathcal{M}$. For the edge states we can use the expression~(\ref{eqn:edge-state}), where $\lambda_- \approx \frac{|\mathcal{M}|\sqrt{\mathcal{B}^2-\mathcal{D}^2}}{2\mathcal{A}\mathcal{B}}$, $\lambda_+\approx \frac{\mathcal{A}}{\sqrt{\mathcal{B}^2-\mathcal{D}^2}}\gg \lambda_-$.

The bulk eigen-state of Hamiltonian~\ref{eqn:BHZ-Ham} for energy $\varepsilon>|\mathcal{M}|$ is a sum of a right-moving wave, a left-moving wave and a term localized in the vicinity of the boundary:
\begin{align}
\psi^{(bulk)}_{s,k_y} &\propto \psi^L + t_R \psi^R + t_d \psi^d\\
\psi^{L} &= e^{-ik_xx}\left(-\frac{Ask_x}{M}\left|E_1,s\right\rangle + \left|H_1,s\right\rangle \right)\\
\psi^{R} &= e^{ik_xx}\left(\frac{Ask_x}{M}\left|E_1,s\right\rangle + \left|H_1,s\right\rangle \right)\\
\psi^{d} &= e^{-\kappa x}\left(\frac{iAs}{B\kappa}\left|E_1,s\right\rangle + \left|H_1,s\right\rangle \right),
\end{align}
where $\kappa \approx \lambda_+$, $k_x = \frac{\sqrt{\varepsilon^2-\mathcal{M}^2}}{\mathcal{A}}\ll \kappa,\lambda_-$. Amplitudes $t_R$ and $t_d$ can be obtained using the zero-boundary condition $\psi^{(bulk)}(x=0)=0$. It yields
\begin{align}
&t_R = \frac{\mathcal{B}k_x\kappa + i\mathcal{M}}{\mathcal{B}k_x\kappa - i\mathcal{M}}\approx -1,\quad
&t_d = -\frac{2\mathcal{B}k_x \kappa}{\mathcal{B}k_x\kappa - i\mathcal{M}}\approx -2\frac{\mathcal{B} ik_x \kappa}{\mathcal{M}}
\end{align}

Now the matrix elements can be calculated straightforwardly. First, we find the overlap integrals defined in the section~\ref{sec:transitions}:
\begin{align}
&F = -\frac{2ik_x}{\lambda_-^2} \frac{1}{\sqrt{2L_x}},\quad
&G = -\frac{2\mathcal{A}k_x}{\mathcal{M} \lambda_- } \frac{1}{\sqrt{2L_x}}
\end{align}
The matrix elements of the transitions can be calculated as
\begin{align}
&|w_\uparrow|^2 =|G|^2 \frac{|\mathcal{B}-\mathcal{D}|}{4|\mathcal{B}|\lambda_-} = \frac{\mathcal{A}^2 k_x^2}{2|\mathcal{M}|^2\lambda_-^3 L_x}\frac{|\mathcal{B}-\mathcal{D}|}{|\mathcal{B}|}\\
&|w_\downarrow|^2=|F|^2 \frac{1}{2\lambda_-} = \frac{2k_x^2}{\lambda_-^5 L_x}
\end{align}

\section{Matrix elements for the light incident at an arbitrary angle\label{app:angle}}
In this section we derive the matrix elements~(\ref{eqn:angle}) for light incident in the direction $\mathbf{n}^{\theta,\phi}$. We can take an auxiliary orthonormal basis
\begin{align}
\mathbf{e}_{x}^{\theta,\phi} &= \mathbf{e}_x \cos\phi\cos\theta + \mathbf{e}_y
\sin\phi\cos\theta-\mathbf{e}_z\sin\theta\\
\mathbf{e}_{y}^{\theta,\phi} &= -\mathbf{e}_x \sin\phi + \mathbf{e}_y\cos\phi\\
\mathbf{e}_{z}^{\theta,\phi} &= \mathbf{n}^{\theta,\phi}=
\mathbf{e}_x \cos\phi\sin\theta + \mathbf{e}_y
\sin\phi\sin\theta+\mathbf{e}_z\cos\theta
\end{align}
The vector potential reads
\begin{align}
&\mathbf{A}^{\theta,\phi}=\mathbf{e}_x^{\theta,\phi} A_0\cos\omega_0t -\mathbf{e}_y^{\theta,\phi} A_0\sin\omega_0t\\
&A_x^{\theta,\phi}= A_0\left(\cos\phi\cos\theta\cos\omega_0t + \sin\phi\sin\omega_0t\right)\\
&A_y^{\theta,\phi}= A_0\left(\sin\phi\cos\theta\cos\omega_0t - \cos\phi\sin\omega_0t\right)\\
&A_z^{\theta,\phi}= -A_0\sin\theta\cos\omega_0t
\end{align}
where the amplitude $A_0=\frac{\sqrt{4\pi \mathcal{W}}}{n_{r}\omega_0}$ (cf.~Eq.~\ref{eqn:vector-potential}). Since only in-plane components of the vector potential yield the optical transitions, the case of arbitrary direction of light is equivalent to the case of elliptical polarization with $\sin\theta$ standing for the eccentricity.

The light-matter interaction Hamiltonian~(\ref{eqn:light-matter-Ham}) can be now rewritten as
\begin{multline}
\hat{H}_{e-A}^{\theta,\phi} = \frac{A_0}{2}e^{i\omega_0t}
\left[ \hat{H}_1 e^{-i\varphi}(\cos\theta +1)+\right.\\\left.
\hat{H}_1^\dag e^{i\varphi}(\cos\theta -1)\right] + H.c.
\end{multline}
Time-reversal symmetry allows us to relate the matrix elements of $\hat{H}_1$ and $\hat{H}_1^\dag$
\begin{align}
\left\langle \psi_{edge,s} \middle| \hat{H}_1^\dag \middle| \psi_{bulk,s} \right\rangle=
-\left\langle \psi_{edge,-s} \middle| \hat{H}_1 \middle| \psi_{bulk,-s} \right\rangle^*
\end{align}
This imply the relation between the matrix elements for different directions of the light
\begin{align}
w_s^{\theta,\phi} = \frac{w_se^{-i\phi}(\cos\theta + 1) + w_{-s}^*e^{i\phi}(1-\cos\theta)}{2}
\end{align}

\section{Derivation of kinetic equation \label{sec:ke-appendix}}

In order to derive kinetic equation it is convenient to write a
second-quantized Hamiltonian by expanding the field operator in eigen
basis of $\hat H_{BHZ}$: $\hat \psi = \sum\limits_{k_y} \hat a_{k_y}
\psi^{(e)}_{k_y} + \sum\limits_{k_x,k_y} \left[\hat c_{k_x,k_y}
  \psi^{(c)}_{k_x}+\hat v_{k_x,k_y} \psi^{(v)}_{k_x} \right]$, where
$\psi^{(e)}$ is the $4\times 1$ wavefunction of the
edge mode, $\psi^{(c)}$, $\psi^{(v)}$ are $4\times 1$ wavefunctions in
conduction and valence bands correspondingly. In this basis the
Hamiltonian reads
\begin{align}
\begin{split}
  \hat H_{BHZ} &=\; \sum\limits_{s,k_y} \varepsilon_{s,k_y} \hat
  a_{s,k_y}^\dag\hat a_{s,k_y} + \sum\limits_{s,k_x,k_y}
  \epsilon^{(c)}_{k_x,k_y} \hat c_{s,k_y}^\dag\hat c_{s,k_y}+\\&
  \sum\limits_{s,k_x,k_y} \epsilon^{(v)}_{k_x,k_y} \hat
  v_{s,k_x,k_y}^\dag\hat v_{s,k_x,k_y},
  \end{split}
  \label{eqn:BHZ-sec-quantized}  
  \\
  \begin{split}
  \hat H_{e-A} &=\sum\limits_{s,k_x,k_y}\left[w_{s}(\hat{{A}}_x -
    i\hat{{A}}_y) +\right.\\&\left. w_{-s}(\hat{{A}}_x + i\hat{{A}}_y) \right]\hat
  a_{s,k_y}^{\dag}\hat c_{s,k_x,k_y} + H.c.
    \end{split}    
  \label{eqn:int-sec-quantized}
\end{align}
where $\epsilon^{(c)}_{k_x,k_y}$, $\epsilon^{(v)}_{k_x,k_y}$ are
the energies in the conduction and valence bands correspondingly. Here we
disregard transitions between the edge states and the valence band as
the frequency of illumination $\omega_0$ is assumed to be much smaller
than the difference between the Fermi level and the top of the valence
band.

The correlations of quantum fluctuations $\delta \hat{\mathbf{A}}$ are
given by the Green's functions $D^R_{\alpha \beta}(1,1') =
-i\theta(t-t')\left\langle\left[ \delta \hat{{A}}_\alpha(1),
    \delta\hat{{A}}_\beta(1')\right] \right\rangle$, $D^A_{\alpha
  \beta}(1,1') = i\theta(t'-t)\left\langle\left[ \delta
    \hat{{A}}_\alpha(1), \delta\hat{{A}}_\beta(1')\right]
\right\rangle$, $D^K = -i\left\langle\left\{ \delta
    \hat{{A}}_\alpha(1), \delta\hat{{A}}_\beta(1')
  \right\}\right\rangle$. At zero temperature these Green's functions
are of the form
\begin{align}
  D^{R(A)}_{\alpha
    \beta}(\omega,k)&=\;\frac{4\pi}{(n_{r}\omega)^2}\frac{(n_{r}\omega)^2\delta_{\alpha\beta}-k_\alpha
    k_\beta}{\epsilon\omega^2-|k|^2 \pm i0} ,\\\quad D^K &=\;
  (D^R-D^A)\mathrm{sign}\;\omega
\end{align}
However, due to the form of the
Hamiltonian~(\ref{eqn:int-sec-quantized}) it is convenient to
introduce axillary scalar fields $$\hat{\tilde{A}}_s =
w_{s}(\hat{{A}}_x - i\hat{{A}}_y) + w_{-s}(\hat{{A}}_x +
i\hat{{A}}_y).$$ Then the Green's function for these fields
$\tilde{D}^R_{s}(1,1') = -i\theta(t-t')\left\langle\left[ \delta
    \hat{{\tilde{A}}}_s(1), \delta\hat{\tilde{A}}^\dag_s(1')\right]
\right\rangle$, $\tilde{D}^A_{s}(1,1') =
i\theta(t'-t)\left\langle\left[ \delta \hat{\tilde{A}}_s(1), \delta
    \hat{{\tilde{A}}}^\dag_s(1')\right] \right\rangle$, $\tilde{D}^K_s =
-i\left\langle\left\{ \delta \hat{{\tilde{A}}}_s(1),
    \delta\hat{{\tilde{A}}}_\beta(1') \right\}\right\rangle$ take up the form
\begin{multline}
  \tilde{D}^{R(A,K)}_s = D_{xx}|w_s+w_{-s}|^2 + D_{yy}|w_s-w_{-s}|^2
  -\\i(D_{xy}+D_{yx})(w_sw_{-s}^* - w_{-s}w_s^*)
\end{multline}

We treat the interaction between electromagnetic field
$\hat{\mathcal{A}}$ and electrons given
by~(\ref{eqn:int-sec-quantized}) as a perturbation, and use the
Keldysh perturbation theory in order to derive the retarded, advanced
and Keldysh Green's functions $G^{R,A,K}_{edge}$ of electrons in the
edge states.  The Dyson equations reads
\begin{align}
\begin{split}
  &(i\partial_t - \varepsilon_{s,k_y})G^{R(A)}_{edge,s}(t,t') - \int dt'' \Sigma^{R(A)}_s(t,t'')G^{R(A)}_{edge,s}(t'',t') =\\& \delta(t-t') 
\end{split}
\\
\begin{split}
  &(i\partial_t - \varepsilon_{s,k_y})G^{K}_{edge,s}(t,t') -\\& \int dt''
  \left[\Sigma^{R}_s(t,t'')G^{K}_{edge,s}(t'',t') +
    \Sigma^{K}_s(t,t'')G^{A}_{edge,s}(t'',t') \right] = 0
    \end{split}
  \label{eqn:left-Dyson}\\
  \begin{split}
  &\left(-i\partial_{t'}-\varepsilon_{s,k_y}\right)G^K_{edge,s}(t,t')
  -\\& \int dt'' \left[G^{R}_{edge,s}(t,t'')\Sigma^{K}_{s}(t'',t') +
    G^{K}_{edge,s}(t,t'')\Sigma^{A}_{s}(t'',t') \right] = 0
    \end{split}
  \label{eqn:right-Dyson}
\end{align}

The self-energy $\Sigma$ can be represented as the sum of a classical
contribution $\Sigma_{ind}$ a contribution $\Sigma_{sp}$ due to
quantum fluctuations $\delta \hat{\mathcal{A}}$, which is responsible for
spontaneous transitions
\begin{align}
  &\Sigma =\; \Sigma_{ind} + \Sigma_{sp},\\
  &\Sigma^{R(A,K)}_{ind,s} =\; \sum\limits_{s,k_x}
  \langle\tilde{\mathcal{A}_s}(t)\rangle
  G^{R(A,K)}_{c,s}(k_x,t,t')\langle\tilde{\mathcal{A}_s^*}(t')\rangle,
 \\
  &\Sigma_{sp,s}^R(t-t',y-y') =\; \frac{i}{2}\left(
    \tilde{D}^R_sG^K_{c,s} + \tilde{D}^K_sG^R_{c,s} \right)
  \label{eqn:Sigma-sp-R}\\
  &\Sigma_{sp,s}^K(t-t',y-y') =\; \frac{i}{2}\left(
    \tilde{D}^K_sG^K_{c,s} + \tilde{D}^R_sG^R_{c,s} +
    \tilde{D}^A_sG^A_{c,s} \right)
\end{align}

Since the classical value of electromagnetic field $\langle \mathbf{A}
\rangle$ depends on time the self-energy $\Sigma_{ind}$ depend both
on the sum of the times $t_s=t+t'$ and their difference $t_a=t-t'$. However,
the dependence on the sum of the times describes the motion of electrons in
high-frequency electromagnetic field, and we disregard it leaving only
the dependence on the time difference, which is responsible for
induced transitions. After performing the Fourier transform over the
difference time we obtain
\begin{multline}
  \Sigma_{ind,s}^{R(A,K)}(\varepsilon, k_y) = \frac{4\pi\mathcal{W}}{(n_{r}\omega_0)^2}\sum\limits_{k_x}|w_s|^2 G^{R(A,K)}_{c,s}(\varepsilon+\omega_0,k_y)+\\|w_{-s}|^2G^{R(A,K)}_{c,s}(\varepsilon-\omega_0,k_y)
  \label{eqn:Sigma-ind}
\end{multline}
The kinetic equation for the distribution function
$n_s(\varepsilon)$ of electrons in the edge state may be
derived in the standard way by subtracting the Dyson equations with self-energy operators acting from the left~(\ref{eqn:left-Dyson}) and from the
right~(\ref{eqn:right-Dyson}), integrating over $k_y$, and using the
relation $G^K=\left(G^R-G^A\right)\left[1-2n_s(\varepsilon)
\right]$:
\begin{align}
  2\partial_t n(\varepsilon) =
  i\left[1-2n(\varepsilon)\right]
  \left(\Sigma^R(k_y) - \Sigma^A(k_y) \right) - i\Sigma^K(k_y)
  \label{eqn:kinetic-general}
\end{align}

Using Eqs.~(\ref{eqn:Sigma-sp-R})~--(\ref{eqn:Sigma-ind}) and the
expression for the Green's functions in the conduction band
$G^{R(A)}_{c,s}=1/\left( \varepsilon - \epsilon_{k_x,k_y} \pm i0
\right)$, $G^K=\left(G^R-G^A\middle)\middle[1- 2N_s\right]$, where
$N_s(\varepsilon)$ is the distribution function of electrons in the
conduction band, the summation over $k_x$ in the self-energy operators
is straightforward:
\begin{align}
  &\Sigma^R_{ind}(\varepsilon,k_y)-\Sigma^A_{ind}(\varepsilon,k_y) =\; -2\pi i\frac{4\pi\mathcal{W}}{(n_{r}\omega_0)^2} |w_s|^2 \tilde{\nu}_{c,k_y}(\varepsilon+\omega_0)\label{eqn:Gamma-ind}\\
  \begin{split}
  &\Sigma^K_{ind}(\varepsilon,k_y) =\;\\& -2\pi i\frac{4\pi\mathcal{W}}{(n_{r}\omega_0)^2}
  |w_s|^2
  \tilde{\nu}_{c,k_y}(\varepsilon+\omega_0)\left[1-2N_s(\varepsilon+\omega_0)\right]
  \end{split}
  \\
  \begin{split}
    & \Sigma^R_{sp}(\varepsilon,k_y)-\Sigma^A_{sp}(\varepsilon,k_y) =\;\\& -\frac{8 i}{3}\int \left(|w_s|^2+|w_{-s}|^2 \right) \tilde{\nu}_{c,k_y}(\varepsilon+\omega)\; N(\varepsilon+\omega)\omega n_{r}\; d\omega
    \end{split} \label{eqn:Gamma-sp}\\
    \begin{split}
    &\Sigma^K_{sp}(\varepsilon,k_y) =\\&-\frac{8 i}{3}\int \left(|w_s|^2+|w_{-s}|^2 \right) \tilde{\nu}_{c,k_y}(\varepsilon+\omega)\; N(\varepsilon+\omega)\omega n_{r}\; d\omega
    \end{split}
\end{align}

The final kinetic equation can be obtained by substituting the self-energy operators into~(\ref{eqn:kinetic-general}):

\begin{align}
\begin{split}
\frac{dn_s(\varepsilon)}{dt} &=\;
-\frac{n_s(\varepsilon)-N_s(\varepsilon+\omega_0)}{\tau_{ind,s}(\varepsilon,\varepsilon+\omega_0)}\mathcal{W}
+\\&\int
\frac{N_s(\varepsilon+\omega)\left[1-n_s(\varepsilon)\right]}{\tau_{sp}(\varepsilon+\omega,\varepsilon)}d\omega
\end{split}
 \\
\tau_{ind,s}^{-1} &=\;
8\pi^2\tilde{\nu}_{C,k_y}(\varepsilon+\omega_0)\frac{|w_s|^2}{n_{r}^2\omega_0^2},\\
 \tau_{sp}^{-1} &=\; \frac{8}{3}\left(|w_s|^2+|w_{-s}|^2 \right)\omega\tilde{\nu}_{C,k_y}(\varepsilon+\omega)n_{r}
\end{align}

\section{EMF calculation\label{sec:emf-appendix}}
From the kinetic equation~(\ref{eqn:kinetic-lead}), one can express the distribution function $f_s$ and its angle-averaged value $\left\langle f_s \right\rangle_\varphi$ via the tunneling source $\Pi_s$ defined in~(\ref{eqn:tunneling-source}):
\begin{align}
\left\langle f_s \right\rangle_\varphi = \left(\left\langle \frac{1}{i\mathbf{v} \mathbf{k}_\parallel + \tau^{-1} }
\right\rangle_\varphi -\tau^{-1}\right)^{-1}
\left( -\Pi_s+|e|\mathbf{v}\mathbf{E}\frac{\partial f_{0}}{\partial \varepsilon} \right),
\end{align}
where $\mathbf{k} = \left(\mathbf{k}_\parallel, k_z \right)$ is a parameter of spatial Fourier transform. 

The quasi-neutrality condition implies 
\begin{align}
\int \left\langle \frac{\mathbf{v}|e|\mathbf{E}\frac{\partial f_{0}}{\partial \varepsilon} }{i\mathbf{v}\mathbf{k}_\parallel + \tau^{-1} } \right\rangle_\varphi \; d\xi 
=
\sum\limits_s \int \left\langle \frac{\Pi_s}{i\mathbf{v}\mathbf{k}_\parallel + \tau^{-1} }
\right\rangle_\varphi \;d\xi
\end{align}
Using the relation $\left\langle \frac{1}{i\mathbf{v}\mathbf{k}_\parallel + \tau^{-1} }
\right\rangle_\varphi = \frac{\tau}{\sqrt{1+k_\parallel^2 l_\tau^2}}$, and $\left\langle \frac{\mathbf{v}\mathbf{E}}{i\mathbf{v}\mathbf{k}_\parallel + \tau^{-1} }
\right\rangle_\varphi = -\frac{i\mathbf{k}_\parallel\mathbf{E}}{k_\parallel^2}\left(1- \frac{1}{\sqrt{1+k_\parallel^2 l_\tau^2}}\right)$ we obtain the final equation for the electric field
\begin{align}
|e|\mathbf{k}_\parallel \mathbf{E} = \frac{i}{2}\sum\limits_s\int \left\langle \frac{\Pi_s}{i\mathbf{v}\mathbf{k}_\parallel + \tau^{-1} }
\right\rangle_\varphi \;d\xi \frac{k_\parallel^2\sqrt{1+k_\parallel^2 l_\tau^2}}{\sqrt{1+k_\parallel^2 l_\tau^2}-1}
\end{align}
The electric field averaged over the lateral direction $x$ corresponds to the $k_x=0$ component:
\begin{multline}
\left\langle E_y\right\rangle_x (y) =\\ \frac{i}{2L_x}\sum\limits_s\int \left\langle\tilde{\Pi}_s \frac{e^{ik_y y}}{ivk_y\cos\varphi + \tau^{-1} }
\right\rangle_\varphi \times \\ \frac{2\sin\frac{k_y\Lambda}{2}\sqrt{1+k_y^2 l_\tau^2}}{\sqrt{1+k_y^2 l_\tau^2}-1}\;d\xi\frac{dk_y}{2\pi},
\end{multline}
where $\tilde{\Pi}_s = 2\pi|\mathcal{T}|^2  \delta(x) \delta_\Gamma\left(\varepsilon^{}_{p_y,s} -
\varepsilon^{(2D)}_{\mathbf{p},s}
\right)\left[f_s-n_s\right] $
We perform averaging over $\varphi$, assuming that the angles with $\cos \varphi$ close to  $s \frac{\varepsilon^{2D} - \varepsilon_0}{p_Fv_{TI}}$ give the main contribution. After summation over spin index and integrating over $\xi = v_F(|p|-p_F)$ we obtain
\begin{multline}
\left\langle E_y\right\rangle_x (y) =\\ \frac{\pi|\mathcal{T}|^2\delta \mu}{L_xp_F v_{TI}\tan \varphi_0} \int  \frac{vk_y e^{ik_y y}}{v^2k_y^2\cos^2\varphi_0 + \tau^{-2} }
\frac{2\sin\frac{k_y\Lambda}{2}\sqrt{1+k_y^2 l_\tau^2}}{\sqrt{1+k_y^2 l_\tau^2}-1}\;\frac{dk_y}{2\pi}
\end{multline}
The EMF can be obtained as the integral of the mean electric field $\mathcal{E} = \int\limits_{-L_y/2}^{L_y/2} \langle E_y \rangle_x \; dy$:
\begin{align}
 \mathcal{E} = 
 \frac{2\pi v|\mathcal{T}|^2\delta \mu}{L_x\tan \varphi_0} \int  \frac{\sin \frac{k_y L_y}{2}}{v^2k_y^2\cos^2\varphi_0 + \tau^{-2} }
 \frac{2\sin\frac{k_y\Lambda}{2}\sqrt{1+k_y^2 l_\tau^2}}{\sqrt{1+k_y^2 l_\tau^2}-1}\;d\xi\frac{dk_y}{2\pi}
\end{align}
The integration can be easily performed if $L_y, \Lambda \gg l_\tau$, i.e. when it is dominated by smal values of $k_y$
\begin{align}
|e|\mathcal{E}  =  -\frac{|\mathcal{T}|^2}{v_F}\frac{2(\varepsilon_F^{(TI)}-\varepsilon_{D})}{\sqrt{\left(p_Fv_{TI}\right)^2-(\varepsilon_F - \varepsilon_{D})^2}} \frac{\Lambda}{L_x}\frac{\delta\mu}{v_{TI}p_F}.
\end{align}


\begin{thebibliography}{34}%
	\makeatletter
	\providecommand \@ifxundefined [1]{%
		\@ifx{#1\undefined}
	}%
	\providecommand \@ifnum [1]{%
		\ifnum #1\expandafter \@firstoftwo
		\else \expandafter \@secondoftwo
		\fi
	}%
	\providecommand \@ifx [1]{%
		\ifx #1\expandafter \@firstoftwo
		\else \expandafter \@secondoftwo
		\fi
	}%
	\providecommand \natexlab [1]{#1}%
	\providecommand \enquote  [1]{``#1''}%
	\providecommand \bibnamefont  [1]{#1}%
	\providecommand \bibfnamefont [1]{#1}%
	\providecommand \citenamefont [1]{#1}%
	\providecommand \href@noop [0]{\@secondoftwo}%
	\providecommand \href [0]{\begingroup \@sanitize@url \@href}%
	\providecommand \@href[1]{\@@startlink{#1}\@@href}%
	\providecommand \@@href[1]{\endgroup#1\@@endlink}%
	\providecommand \@sanitize@url [0]{\catcode `\\12\catcode `\$12\catcode
		`\&12\catcode `\#12\catcode `\^12\catcode `\_12\catcode `\%12\relax}%
	\providecommand \@@startlink[1]{}%
	\providecommand \@@endlink[0]{}%
	\providecommand \url  [0]{\begingroup\@sanitize@url \@url }%
	\providecommand \@url [1]{\endgroup\@href {#1}{\urlprefix }}%
	\providecommand \urlprefix  [0]{URL }%
	\providecommand \Eprint [0]{\href }%
	\providecommand \doibase [0]{http://dx.doi.org/}%
	\providecommand \selectlanguage [0]{\@gobble}%
	\providecommand \bibinfo  [0]{\@secondoftwo}%
	\providecommand \bibfield  [0]{\@secondoftwo}%
	\providecommand \translation [1]{[#1]}%
	\providecommand \BibitemOpen [0]{}%
	\providecommand \bibitemStop [0]{}%
	\providecommand \bibitemNoStop [0]{.\EOS\space}%
	\providecommand \EOS [0]{\spacefactor3000\relax}%
	\providecommand \BibitemShut  [1]{\csname bibitem#1\endcsname}%
	\let\auto@bib@innerbib\@empty
	\bibitem [{\citenamefont {K{\"o}nig}\ \emph {et~al.}(2008)\citenamefont
		{K{\"o}nig}, \citenamefont {Buhmann}, \citenamefont {Molenkamp},
		\citenamefont {Hughes}, \citenamefont {Liu}, \citenamefont {Qi},\ and\
		\citenamefont {Zhang}}]{KonigJPhysSocJpn2008}%
	\BibitemOpen
	\bibfield  {author} {\bibinfo {author} {\bibfnamefont {M.}~\bibnamefont
			{K{\"o}nig}}, \bibinfo {author} {\bibfnamefont {H.}~\bibnamefont {Buhmann}},
		\bibinfo {author} {\bibfnamefont {L.~W.}\ \bibnamefont {Molenkamp}}, \bibinfo
		{author} {\bibfnamefont {T.}~\bibnamefont {Hughes}}, \bibinfo {author}
		{\bibfnamefont {C.-X.}\ \bibnamefont {Liu}}, \bibinfo {author} {\bibfnamefont
			{X.-L.}\ \bibnamefont {Qi}}, \ and\ \bibinfo {author} {\bibfnamefont {S.-C.}\
			\bibnamefont {Zhang}},\ }\href@noop {} {\bibfield  {journal} {\bibinfo
			{journal} {J. Phys. Soc. Jpn.}\ }\textbf {\bibinfo {volume} {77}},\ \bibinfo
		{pages} {031007} (\bibinfo {year} {2008})}\BibitemShut {NoStop}%
	\bibitem [{\citenamefont {Hasan}\ and\ \citenamefont
		{Kane}(2010)}]{HasanKane2010colloqium}%
	\BibitemOpen
	\bibfield  {author} {\bibinfo {author} {\bibfnamefont {M.~Z.}\ \bibnamefont
			{Hasan}}\ and\ \bibinfo {author} {\bibfnamefont {C.~L.}\ \bibnamefont
			{Kane}},\ }\href@noop {} {\bibfield  {journal} {\bibinfo  {journal} {Rev.
				Mod. Phys.}\ }\textbf {\bibinfo {volume} {82}},\ \bibinfo {pages} {3045}
		(\bibinfo {year} {2010})}\BibitemShut {NoStop}%
	\bibitem [{\citenamefont {Qi}\ and\ \citenamefont
		{Zhang}(2011)}]{QiZhangRevModPhys2011}%
	\BibitemOpen
	\bibfield  {author} {\bibinfo {author} {\bibfnamefont {X.-L.}\ \bibnamefont
			{Qi}}\ and\ \bibinfo {author} {\bibfnamefont {S.-C.}\ \bibnamefont {Zhang}},\
	}\href@noop {} {\bibfield  {journal} {\bibinfo  {journal} {Reviews of Modern
			Physics}\ }\textbf {\bibinfo {volume} {83}},\ \bibinfo {pages} {1057}
	(\bibinfo {year} {2011})}\BibitemShut {NoStop}%
\bibitem [{\citenamefont {Muniz}\ and\ \citenamefont
	{Sipe}(2014)}]{MunizSipePRB2014}%
\BibitemOpen
\bibfield  {author} {\bibinfo {author} {\bibfnamefont {R.~A.}\ \bibnamefont
		{Muniz}}\ and\ \bibinfo {author} {\bibfnamefont {J.~E.}~\bibnamefont {Sipe}},\
}\href@noop {} {\bibfield  {journal} {\bibinfo  {journal} {Physical Review
		B}\ }\textbf {\bibinfo {volume} {89}},\ \bibinfo {pages} {205113} (\bibinfo
{year} {2014})}\BibitemShut {NoStop}%
\bibitem [{\citenamefont {McIver}\ \emph {et~al.}(2012)\citenamefont {McIver},
	\citenamefont {Hsieh}, \citenamefont {Steinberg}, \citenamefont
	{Jarillo-Herrero},\ and\ \citenamefont {Gedik}}]{MciverNature2012}%
\BibitemOpen
\bibfield  {author} {\bibinfo {author} {\bibfnamefont {J.}~\bibnamefont
		{McIver}}, \bibinfo {author} {\bibfnamefont {D.}~\bibnamefont {Hsieh}},
	\bibinfo {author} {\bibfnamefont {H.}~\bibnamefont {Steinberg}}, \bibinfo
	{author} {\bibfnamefont {P.}~\bibnamefont {Jarillo-Herrero}}, \ and\ \bibinfo
	{author} {\bibfnamefont {N.}~\bibnamefont {Gedik}},\ }\href@noop {}
{\bibfield  {journal} {\bibinfo  {journal} {Nature nanotechnology}\ }\textbf
	{\bibinfo {volume} {7}},\ \bibinfo {pages} {96} (\bibinfo {year}
	{2012})}\BibitemShut {NoStop}%
\bibitem [{\citenamefont {Misawa}\ \emph {et~al.}(2011)\citenamefont {Misawa},
	\citenamefont {Yokoyama},\ and\ \citenamefont {Murakami}}]{MisawaPRB2011}%
\BibitemOpen
\bibfield  {author} {\bibinfo {author} {\bibfnamefont {T.}~\bibnamefont
		{Misawa}}, \bibinfo {author} {\bibfnamefont {T.}~\bibnamefont {Yokoyama}}, \
	and\ \bibinfo {author} {\bibfnamefont {S.}~\bibnamefont {Murakami}},\
}\href@noop {} {\bibfield  {journal} {\bibinfo  {journal} {Physical Review
		B}\ }\textbf {\bibinfo {volume} {84}},\ \bibinfo {pages} {165407} (\bibinfo
{year} {2011})}\BibitemShut {NoStop}%
\bibitem [{\citenamefont {Junck}\ \emph {et~al.}(2013)\citenamefont {Junck},
	\citenamefont {Refael},\ and\ \citenamefont {von Oppen}}]{JunckPRB2013}%
\BibitemOpen
\bibfield  {author} {\bibinfo {author} {\bibfnamefont {A.}~\bibnamefont
		{Junck}}, \bibinfo {author} {\bibfnamefont {G.}~\bibnamefont {Refael}}, \
	and\ \bibinfo {author} {\bibfnamefont {F.}~\bibnamefont {von Oppen}},\
}\href@noop {} {\bibfield  {journal} {\bibinfo  {journal} {Physical Review
		B}\ }\textbf {\bibinfo {volume} {88}},\ \bibinfo {pages} {075144} (\bibinfo
{year} {2013})}\BibitemShut {NoStop}%
\bibitem [{\citenamefont {Hosur}(2011)}]{HosurPRB2011}%
\BibitemOpen
\bibfield  {author} {\bibinfo {author} {\bibfnamefont {P.}~\bibnamefont
		{Hosur}},\ }\href@noop {} {\bibfield  {journal} {\bibinfo  {journal}
		{Physical Review B}\ }\textbf {\bibinfo {volume} {83}},\ \bibinfo {pages}
	{035309} (\bibinfo {year} {2011})}\BibitemShut {NoStop}%
\bibitem [{\citenamefont {Jozwiak}\ \emph {et~al.}(2013)\citenamefont
	{Jozwiak}, \citenamefont {Park}, \citenamefont {Gotlieb}, \citenamefont
	{Hwang}, \citenamefont {Lee}, \citenamefont {Louie}, \citenamefont
	{Denlinger}, \citenamefont {Rotundu}, \citenamefont {Birgeneau},
	\citenamefont {Hussain} \emph {et~al.}}]{JozwiakNatPhys2013}%
\BibitemOpen
\bibfield  {author} {\bibinfo {author} {\bibfnamefont {C.}~\bibnamefont
		{Jozwiak}}, \bibinfo {author} {\bibfnamefont {C.-H.}\ \bibnamefont {Park}},
	\bibinfo {author} {\bibfnamefont {K.}~\bibnamefont {Gotlieb}}, \bibinfo
	{author} {\bibfnamefont {C.}~\bibnamefont {Hwang}}, \bibinfo {author}
	{\bibfnamefont {D.-H.}\ \bibnamefont {Lee}}, \bibinfo {author} {\bibfnamefont
		{S.~G.}\ \bibnamefont {Louie}}, \bibinfo {author} {\bibfnamefont {J.~D.}\
		\bibnamefont {Denlinger}}, \bibinfo {author} {\bibfnamefont {C.~R.}\
		\bibnamefont {Rotundu}}, \bibinfo {author} {\bibfnamefont {R.~J.}\
		\bibnamefont {Birgeneau}}, \bibinfo {author} {\bibfnamefont {Z.}~\bibnamefont
		{Hussain}},  \emph {et~al.},\ }\href@noop {} {\bibfield  {journal} {\bibinfo
		{journal} {Nature Physics}\ }\textbf {\bibinfo {volume} {9}},\ \bibinfo
	{pages} {293} (\bibinfo {year} {2013})}\BibitemShut {NoStop}%
\bibitem [{\citenamefont {Zhang}\ \emph {et~al.}(2010)\citenamefont {Zhang},
	\citenamefont {Wang},\ and\ \citenamefont {Zhang}}]{ZhangPRB2010}%
\BibitemOpen
\bibfield  {author} {\bibinfo {author} {\bibfnamefont {X.}~\bibnamefont
		{Zhang}}, \bibinfo {author} {\bibfnamefont {J.}~\bibnamefont {Wang}}, \ and\
	\bibinfo {author} {\bibfnamefont {S.-C.}\ \bibnamefont {Zhang}},\ }\href@noop
{} {\bibfield  {journal} {\bibinfo  {journal} {Physical Review B}\ }\textbf
	{\bibinfo {volume} {82}},\ \bibinfo {pages} {245107} (\bibinfo {year}
	{2010})}\BibitemShut {NoStop}%
\bibitem [{\citenamefont {Park}\ and\ \citenamefont
	{Louie}(2012)}]{ParkLouiePRL2012}%
\BibitemOpen
\bibfield  {author} {\bibinfo {author} {\bibfnamefont {C.-H.}\ \bibnamefont
		{Park}}\ and\ \bibinfo {author} {\bibfnamefont {S.~G.}\ \bibnamefont
		{Louie}},\ }\href@noop {} {\bibfield  {journal} {\bibinfo  {journal}
		{Physical review letters}\ }\textbf {\bibinfo {volume} {109}},\ \bibinfo
	{pages} {097601} (\bibinfo {year} {2012})}\BibitemShut {NoStop}%
\bibitem [{\citenamefont {Wu}\ \emph {et~al.}(2012)\citenamefont {Wu},
	\citenamefont {Zhang}, \citenamefont {Fang},\ and\ \citenamefont
	{Dai}}]{WuPhysE2012}%
\BibitemOpen
\bibfield  {author} {\bibinfo {author} {\bibfnamefont {Q.~S.}\ \bibnamefont
		{Wu}}, \bibinfo {author} {\bibfnamefont {S.~N.}\ \bibnamefont {Zhang}},
	\bibinfo {author} {\bibfnamefont {Z.}~\bibnamefont {Fang}}, \ and\ \bibinfo
	{author} {\bibfnamefont {X.}~\bibnamefont {Dai}},\ }\href@noop {} {\bibfield
	{journal} {\bibinfo  {journal} {Physica E: Low-dimensional Systems and
			Nanostructures}\ }\textbf {\bibinfo {volume} {44}},\ \bibinfo {pages} {895}
	(\bibinfo {year} {2012})}\BibitemShut {NoStop}%
\bibitem [{\citenamefont {Zhu}\ \emph {et~al.}(2014)\citenamefont {Zhu},
	\citenamefont {Veenstra}, \citenamefont {Zhdanovich}, \citenamefont
	{Schneider}, \citenamefont {Okuda}, \citenamefont {Miyamoto}, \citenamefont
	{Zhu}, \citenamefont {Namatame}, \citenamefont {Taniguchi}, \citenamefont
	{Haverkort} \emph {et~al.}}]{ZhuPRL2014}%
\BibitemOpen
\bibfield  {author} {\bibinfo {author} {\bibfnamefont {Z.-H.}\ \bibnamefont
		{Zhu}}, \bibinfo {author} {\bibfnamefont {C.}~\bibnamefont {Veenstra}},
	\bibinfo {author} {\bibfnamefont {S.}~\bibnamefont {Zhdanovich}}, \bibinfo
	{author} {\bibfnamefont {M.}~\bibnamefont {Schneider}}, \bibinfo {author}
	{\bibfnamefont {T.}~\bibnamefont {Okuda}}, \bibinfo {author} {\bibfnamefont
		{K.}~\bibnamefont {Miyamoto}}, \bibinfo {author} {\bibfnamefont {S.-Y.}\
		\bibnamefont {Zhu}}, \bibinfo {author} {\bibfnamefont {H.}~\bibnamefont
		{Namatame}}, \bibinfo {author} {\bibfnamefont {M.}~\bibnamefont {Taniguchi}},
	\bibinfo {author} {\bibfnamefont {M.}~\bibnamefont {Haverkort}},  \emph
	{et~al.},\ }\href@noop {} {\bibfield  {journal} {\bibinfo  {journal}
		{Physical review letters}\ }\textbf {\bibinfo {volume} {112}},\ \bibinfo
	{pages} {076802} (\bibinfo {year} {2014})}\BibitemShut {NoStop}%
\bibitem [{\citenamefont {Li}\ and\ \citenamefont
	{Carbotte}(2013)}]{PRB-88-045414}%
\BibitemOpen
\bibfield  {author} {\bibinfo {author} {\bibfnamefont {Z.}~\bibnamefont
		{Li}}\ and\ \bibinfo {author} {\bibfnamefont {J.~P.}\ \bibnamefont
		{Carbotte}},\ }\href {\doibase 10.1103/PhysRevB.88.045414} {\bibfield
	{journal} {\bibinfo  {journal} {Phys. Rev. B}\ }\textbf {\bibinfo {volume}
		{88}},\ \bibinfo {pages} {045414} (\bibinfo {year} {2013})}\BibitemShut
{NoStop}%
\bibitem [{\citenamefont {Wilson}\ \emph {et~al.}(2014)\citenamefont {Wilson},
	\citenamefont {Efimkin},\ and\ \citenamefont {Galitski}}]{PRB-90-205432}%
\BibitemOpen
\bibfield  {author} {\bibinfo {author} {\bibfnamefont {J.~H.}\ \bibnamefont
		{Wilson}}, \bibinfo {author} {\bibfnamefont {D.~K.}\ \bibnamefont {Efimkin}},
	\ and\ \bibinfo {author} {\bibfnamefont {V.~M.}\ \bibnamefont {Galitski}},\
}\href {\doibase 10.1103/PhysRevB.90.205432} {\bibfield  {journal} {\bibinfo
	{journal} {Phys. Rev. B}\ }\textbf {\bibinfo {volume} {90}},\ \bibinfo
{pages} {205432} (\bibinfo {year} {2014})}\BibitemShut {NoStop}%
\bibitem [{\citenamefont {Bernevig}\ \emph {et~al.}(2006)\citenamefont
	{Bernevig}, \citenamefont {Hughes},\ and\ \citenamefont
	{Zhang}}]{BernevigHughesZhangScience2006}%
\BibitemOpen
\bibfield  {author} {\bibinfo {author} {\bibfnamefont {B.~A.}\ \bibnamefont
		{Bernevig}}, \bibinfo {author} {\bibfnamefont {T.~L.}\ \bibnamefont
		{Hughes}}, \ and\ \bibinfo {author} {\bibfnamefont {S.-C.}\ \bibnamefont
		{Zhang}},\ }\href@noop {} {\bibfield  {journal} {\bibinfo  {journal}
		{Science}\ }\textbf {\bibinfo {volume} {314}},\ \bibinfo {pages} {1757}
	(\bibinfo {year} {2006})}\BibitemShut {NoStop}%
\bibitem [{\citenamefont {Diehl}\ \emph {et~al.}(2009)\citenamefont {Diehl},
	\citenamefont {Shalygin}, \citenamefont {Golub}, \citenamefont {Tarasenko},
	\citenamefont {Danilov}, \citenamefont {Bel'kov}, \citenamefont {Novik},
	\citenamefont {Buhmann}, \citenamefont {Br{\"u}ne}, \citenamefont {Molenkamp}
	\emph {et~al.}}]{DiehlPRB2009}%
\BibitemOpen
\bibfield  {author} {\bibinfo {author} {\bibfnamefont {H.}~\bibnamefont
		{Diehl}}, \bibinfo {author} {\bibfnamefont {V.~A.}\ \bibnamefont {Shalygin}},
	\bibinfo {author} {\bibfnamefont {L.~E.}\ \bibnamefont {Golub}}, \bibinfo
	{author} {\bibfnamefont {S.~A.}\ \bibnamefont {Tarasenko}}, \bibinfo {author}
	{\bibfnamefont {S.~N.}\ \bibnamefont {Danilov}}, \bibinfo {author}
	{\bibfnamefont {V.~V.}\ \bibnamefont {Bel'kov}}, \bibinfo {author}
	{\bibfnamefont {E.~G.}\ \bibnamefont {Novik}}, \bibinfo {author}
	{\bibfnamefont {H.}~\bibnamefont {Buhmann}}, \bibinfo {author} {\bibfnamefont
		{C.}~\bibnamefont {Br{\"u}ne}}, \bibinfo {author} {\bibfnamefont {L.~W.}\
		\bibnamefont {Molenkamp}},  \emph {et~al.},\ }\href@noop {} {\bibfield
	{journal} {\bibinfo  {journal} {Physical Review B}\ }\textbf {\bibinfo
		{volume} {80}},\ \bibinfo {pages} {075311} (\bibinfo {year}
	{2009})}\BibitemShut {NoStop}%
\bibitem [{\citenamefont {Wittmann}\ \emph {et~al.}(2010)\citenamefont
	{Wittmann}, \citenamefont {Danilov}, \citenamefont {Bel'kov}, \citenamefont
	{Tarasenko}, \citenamefont {Novik}, \citenamefont {Buhmann}, \citenamefont
	{Br{\"u}ne}, \citenamefont {Molenkamp}, \citenamefont {Kvon}, \citenamefont
	{Mikhailov} \emph {et~al.}}]{WittmannSemiconductor2010}%
\BibitemOpen
\bibfield  {author} {\bibinfo {author} {\bibfnamefont {B.}~\bibnamefont
		{Wittmann}}, \bibinfo {author} {\bibfnamefont {S.}~\bibnamefont {Danilov}},
	\bibinfo {author} {\bibfnamefont {V.}~\bibnamefont {Bel'kov}}, \bibinfo
	{author} {\bibfnamefont {S.}~\bibnamefont {Tarasenko}}, \bibinfo {author}
	{\bibfnamefont {E.~G.}~\bibnamefont {Novik}}, \bibinfo {author} {\bibfnamefont
		{H.}~\bibnamefont {Buhmann}}, \bibinfo {author} {\bibfnamefont
		{C.}~\bibnamefont {Br{\"u}ne}}, \bibinfo {author} {\bibfnamefont
		{L.~W.}~\bibnamefont {Molenkamp}}, \bibinfo {author} {\bibfnamefont
		{Z.}~\bibnamefont {Kvon}}, \bibinfo {author} {\bibfnamefont {N.}~\bibnamefont
		{Mikhailov}},  \emph {et~al.},\ }\href@noop {} {\bibfield  {journal}
	{\bibinfo  {journal} {Semiconductor Science and Technology}\ }\textbf
	{\bibinfo {volume} {25}},\ \bibinfo {pages} {095005} (\bibinfo {year}
	{2010})}\BibitemShut {NoStop}%
\bibitem [{\citenamefont {Scharf}\ \emph {et~al.}(2015)\citenamefont {Scharf},
	\citenamefont {Matos-Abiague}, \citenamefont {{\v Z}uti{\'c}},\ and\
	\citenamefont {Fabian}}]{arXiv1502.05605}%
\BibitemOpen
\bibfield  {author} {\bibinfo {author} {\bibfnamefont {B.}~\bibnamefont
		{Scharf}}, \bibinfo {author} {\bibfnamefont {A.}~\bibnamefont
		{Matos-Abiague}}, \bibinfo {author} {\bibfnamefont {I.}~\bibnamefont {{\v
				Z}uti{\'c}}}, \ and\ \bibinfo {author} {\bibfnamefont {J.}~\bibnamefont
		{Fabian}},\ }\href@noop {} {\enquote {\bibinfo {title} {Probing topological
			transitions in {H}g{T}e/{C}d{T}e quantum wells by magneto-optical
			measurements},}\ } (\bibinfo {year} {2015}),\ \Eprint
{http://arxiv.org/abs/arXiv:1502.05605} {arXiv:1502.05605} \BibitemShut
{NoStop}%
\bibitem [{\citenamefont {Artemenko}\ and\ \citenamefont
	{Kaladzhyan}(2013)}]{ArtemenkoKaladzhyanJETPLett2013}%
\BibitemOpen
\bibfield  {author} {\bibinfo {author} {\bibfnamefont {S.~N.}\ \bibnamefont
		{Artemenko}}\ and\ \bibinfo {author} {\bibfnamefont {V.}~\bibnamefont
		{Kaladzhyan}},\ }\href@noop {} {\bibfield  {journal} {\bibinfo  {journal}
		{JETP letters}\ }\textbf {\bibinfo {volume} {97}},\ \bibinfo {pages} {82}
	(\bibinfo {year} {2013})}\BibitemShut {NoStop}%
\bibitem [{\citenamefont {Novik}\ \emph {et~al.}(2005)\citenamefont {Novik},
	\citenamefont {Pfeuffer-Jeschke}, \citenamefont {Jungwirth}, \citenamefont
	{Latussek}, \citenamefont {Becker}, \citenamefont {Landwehr}, \citenamefont
	{Buhmann},\ and\ \citenamefont {Molenkamp}}]{NovikPRB2005}%
\BibitemOpen
\bibfield  {author} {\bibinfo {author} {\bibfnamefont {E.~G.}~\bibnamefont
		{Novik}}, \bibinfo {author} {\bibfnamefont {A.}~\bibnamefont
		{Pfeuffer-Jeschke}}, \bibinfo {author} {\bibfnamefont {T.}~\bibnamefont
		{Jungwirth}}, \bibinfo {author} {\bibfnamefont {V.}~\bibnamefont {Latussek}},
	\bibinfo {author} {\bibfnamefont {C.~R.}~\bibnamefont {Becker}}, \bibinfo
	{author} {\bibfnamefont {G.}~\bibnamefont {Landwehr}}, \bibinfo {author}
	{\bibfnamefont {H.}~\bibnamefont {Buhmann}}, \ and\ \bibinfo {author}
	{\bibfnamefont {L.~W.}~\bibnamefont {Molenkamp}},\ }\href@noop {} {\bibfield
	{journal} {\bibinfo  {journal} {Physical Review B}\ }\textbf {\bibinfo
		{volume} {72}},\ \bibinfo {pages} {035321} (\bibinfo {year}
	{2005})}\BibitemShut {NoStop}%
\bibitem [{\citenamefont {Zhang}\ \emph {et~al.}(2009)\citenamefont {Zhang},
	\citenamefont {Liu}, \citenamefont {Qi}, \citenamefont {Dai}, \citenamefont
	{Fang},\ and\ \citenamefont {Zhang}}]{ZhangNaturePhys2009}%
\BibitemOpen
\bibfield  {author} {\bibinfo {author} {\bibfnamefont {H.}~\bibnamefont
		{Zhang}}, \bibinfo {author} {\bibfnamefont {C.-X.}\ \bibnamefont {Liu}},
	\bibinfo {author} {\bibfnamefont {X.-L.}\ \bibnamefont {Qi}}, \bibinfo
	{author} {\bibfnamefont {X.}~\bibnamefont {Dai}}, \bibinfo {author}
	{\bibfnamefont {Z.}~\bibnamefont {Fang}}, \ and\ \bibinfo {author}
	{\bibfnamefont {S.-C.}\ \bibnamefont {Zhang}},\ }\href@noop {} {\bibfield
	{journal} {\bibinfo  {journal} {Nature Physics}\ }\textbf {\bibinfo {volume}
		{5}},\ \bibinfo {pages} {438} (\bibinfo {year} {2009})}\BibitemShut {NoStop}%
\bibitem [{\citenamefont {Erlingsson}\ and\ \citenamefont
	{Egues}(2015)}]{ErlingssonPRB2015}%
\BibitemOpen
\bibfield  {author} {\bibinfo {author} {\bibfnamefont {S.~I.}\ \bibnamefont
		{Erlingsson}}\ and\ \bibinfo {author} {\bibfnamefont {J.~C.}\ \bibnamefont
		{Egues}},\ }\href@noop {} {\bibfield  {journal} {\bibinfo  {journal}
		{Physical Review B}\ }\textbf {\bibinfo {volume} {91}},\ \bibinfo {pages}
	{035312} (\bibinfo {year} {2015})}\BibitemShut {NoStop}%
\bibitem [{\citenamefont {Isaev}\ \emph {et~al.}(2011)\citenamefont {Isaev},
	\citenamefont {Moon},\ and\ \citenamefont {Ortiz}}]{IsaevPRB2011}%
\BibitemOpen
\bibfield  {author} {\bibinfo {author} {\bibfnamefont {L.}~\bibnamefont
		{Isaev}}, \bibinfo {author} {\bibfnamefont {Y.~H.}\ \bibnamefont {Moon}}, \
	and\ \bibinfo {author} {\bibfnamefont {G.}~\bibnamefont {Ortiz}},\ }\href
{\doibase 10.1103/PhysRevB.84.075444} {\bibfield  {journal} {\bibinfo
		{journal} {Phys. Rev. B}\ }\textbf {\bibinfo {volume} {84}},\ \bibinfo
	{pages} {075444} (\bibinfo {year} {2011})}\BibitemShut {NoStop}%
\bibitem [{\citenamefont {Medhi}\ and\ \citenamefont
	{Shenoy}(2012)}]{MedhiJPhysCondMat2012}%
\BibitemOpen
\bibfield  {author} {\bibinfo {author} {\bibfnamefont {A.}~\bibnamefont
		{Medhi}}\ and\ \bibinfo {author} {\bibfnamefont {V.~B.}\ \bibnamefont
		{Shenoy}},\ }\href@noop {} {\bibfield  {journal} {\bibinfo  {journal}
		{Journal of Physics: Condensed Matter}\ }\textbf {\bibinfo {volume} {24}},\
	\bibinfo {pages} {355001} (\bibinfo {year} {2012})}\BibitemShut {NoStop}%
\bibitem [{\citenamefont {Men’shov}\ \emph {et~al.}(2014)\citenamefont
	{Men’shov}, \citenamefont {Tugushev},\ and\ \citenamefont
	{Chulkov}}]{MenshovJETPLett2014}%
\BibitemOpen
\bibfield  {author} {\bibinfo {author} {\bibfnamefont {V.~N.}\ \bibnamefont
		{Men’shov}}, \bibinfo {author} {\bibfnamefont {V.~V.}\ \bibnamefont
		{Tugushev}}, \ and\ \bibinfo {author} {\bibfnamefont {E.~V.}\ \bibnamefont
		{Chulkov}},\ }\href {\doibase 10.1134/S0021364013230082} {\bibfield
	{journal} {\bibinfo  {journal} {JETP Letters}\ }\textbf {\bibinfo {volume}
		{98}},\ \bibinfo {pages} {603} (\bibinfo {year} {2014})}\BibitemShut
{NoStop}%
\bibitem [{\citenamefont {Enaldiev}\ \emph {et~al.}(2015)\citenamefont
	{Enaldiev}, \citenamefont {Zagorodnev},\ and\ \citenamefont
	{Volkov}}]{EnaldievJETPLett2015}%
\BibitemOpen
\bibfield  {author} {\bibinfo {author} {\bibfnamefont {V.~V.}\ \bibnamefont
		{Enaldiev}}, \bibinfo {author} {\bibfnamefont {I.~V.}\ \bibnamefont
		{Zagorodnev}}, \ and\ \bibinfo {author} {\bibfnamefont {V.~A.}\ \bibnamefont
		{Volkov}},\ }\href@noop {} {\bibfield  {journal} {\bibinfo  {journal} {JETP
			Letters}\ }\textbf {\bibinfo {volume} {101}},\ \bibinfo {pages} {89}
	(\bibinfo {year} {2015})}\BibitemShut {NoStop}%
\bibitem [{\citenamefont {Gusev}\ \emph {et~al.}(2014)\citenamefont {Gusev},
	\citenamefont {Kvon}, \citenamefont {Olshanetsky}, \citenamefont {Levin},
	\citenamefont {Krupko}, \citenamefont {Portal}, \citenamefont {Mikhailov},\
	and\ \citenamefont {Dvoretsky}}]{GusevPRB2014}%
\BibitemOpen
\bibfield  {author} {\bibinfo {author} {\bibfnamefont {G.~M.}\ \bibnamefont
		{Gusev}}, \bibinfo {author} {\bibfnamefont {Z.~D.}\ \bibnamefont {Kvon}},
	\bibinfo {author} {\bibfnamefont {E.~B.}\ \bibnamefont {Olshanetsky}},
	\bibinfo {author} {\bibfnamefont {A.~D.}\ \bibnamefont {Levin}}, \bibinfo
	{author} {\bibfnamefont {Y.}~\bibnamefont {Krupko}}, \bibinfo {author}
	{\bibfnamefont {J.~C.}\ \bibnamefont {Portal}}, \bibinfo {author}
	{\bibfnamefont {N.~N.}\ \bibnamefont {Mikhailov}}, \ and\ \bibinfo {author}
	{\bibfnamefont {S.~A.}\ \bibnamefont {Dvoretsky}},\ }\href {\doibase
	10.1103/PhysRevB.89.125305} {\bibfield  {journal} {\bibinfo  {journal} {Phys.
			Rev. B}\ }\textbf {\bibinfo {volume} {89}},\ \bibinfo {pages} {125305}
	(\bibinfo {year} {2014})}\BibitemShut {NoStop}%
\bibitem [{\citenamefont {Maciejko}\ \emph {et~al.}(2009)\citenamefont
	{Maciejko}, \citenamefont {Liu}, \citenamefont {Oreg}, \citenamefont {Qi},
	\citenamefont {Wu},\ and\ \citenamefont {Zhang}}]{MaciejkoPRL2009}%
\BibitemOpen
\bibfield  {author} {\bibinfo {author} {\bibfnamefont {J.}~\bibnamefont
		{Maciejko}}, \bibinfo {author} {\bibfnamefont {C.}~\bibnamefont {Liu}},
	\bibinfo {author} {\bibfnamefont {Y.}~\bibnamefont {Oreg}}, \bibinfo {author}
	{\bibfnamefont {X.-L.}\ \bibnamefont {Qi}}, \bibinfo {author} {\bibfnamefont
		{C.}~\bibnamefont {Wu}}, \ and\ \bibinfo {author} {\bibfnamefont {S.-C.}\
		\bibnamefont {Zhang}},\ }\href {\doibase 10.1103/PhysRevLett.102.256803}
{\bibfield  {journal} {\bibinfo  {journal} {Phys. Rev. Lett.}\ }\textbf
	{\bibinfo {volume} {102}},\ \bibinfo {pages} {256803} (\bibinfo {year}
	{2009})}\BibitemShut {NoStop}%
\bibitem [{\citenamefont {Tanaka}\ \emph {et~al.}(2011)\citenamefont {Tanaka},
	\citenamefont {Furusaki},\ and\ \citenamefont {Matveev}}]{TanakaPRL2011}%
\BibitemOpen
\bibfield  {author} {\bibinfo {author} {\bibfnamefont {Y.}~\bibnamefont
		{Tanaka}}, \bibinfo {author} {\bibfnamefont {A.}~\bibnamefont {Furusaki}}, \
	and\ \bibinfo {author} {\bibfnamefont {K.~A.}\ \bibnamefont {Matveev}},\
}\href {\doibase 10.1103/PhysRevLett.106.236402} {\bibfield  {journal}
{\bibinfo  {journal} {Phys. Rev. Lett.}\ }\textbf {\bibinfo {volume} {106}},\
\bibinfo {pages} {236402} (\bibinfo {year} {2011})}\BibitemShut {NoStop}%
\bibitem [{\citenamefont {Lunde}\ and\ \citenamefont
	{Platero}(2012)}]{LundePRB2012}%
\BibitemOpen
\bibfield  {author} {\bibinfo {author} {\bibfnamefont {A.~M.}\ \bibnamefont
		{Lunde}}\ and\ \bibinfo {author} {\bibfnamefont {G.}~\bibnamefont
		{Platero}},\ }\href {\doibase 10.1103/PhysRevB.86.035112} {\bibfield
	{journal} {\bibinfo  {journal} {Phys. Rev. B}\ }\textbf {\bibinfo {volume}
		{86}},\ \bibinfo {pages} {035112} (\bibinfo {year} {2012})}\BibitemShut
{NoStop}%
\bibitem [{\citenamefont {V\"ayrynen}\ \emph {et~al.}(2014)\citenamefont
	{V\"ayrynen}, \citenamefont {Goldstein}, \citenamefont {Gefen},\ and\
	\citenamefont {Glazman}}]{VayrunenPRB2014}%
\BibitemOpen
\bibfield  {author} {\bibinfo {author} {\bibfnamefont {J.~I.}\ \bibnamefont
		{V\"ayrynen}}, \bibinfo {author} {\bibfnamefont {M.}~\bibnamefont
		{Goldstein}}, \bibinfo {author} {\bibfnamefont {Y.}~\bibnamefont {Gefen}}, \
	and\ \bibinfo {author} {\bibfnamefont {L.~I.}\ \bibnamefont {Glazman}},\
}\href {\doibase 10.1103/PhysRevB.90.115309} {\bibfield  {journal} {\bibinfo
	{journal} {Phys. Rev. B}\ }\textbf {\bibinfo {volume} {90}},\ \bibinfo
{pages} {115309} (\bibinfo {year} {2014})}\BibitemShut {NoStop}%
\bibitem [{\citenamefont {Yariv}(1989)}]{amnon1989quantum}%
\BibitemOpen
\bibfield  {author} {\bibinfo {author} {\bibfnamefont {A.}~\bibnamefont
		{Yariv}},\ }\href {http://books.google.ru/books?id=UTWg1VIkNuMC} {\emph
	{\bibinfo {title} {Quantum electronics}}}\ (\bibinfo  {publisher} {Wiley},\
\bibinfo {year} {1989})\ Chap.~\bibinfo {chapter} {8}\BibitemShut {NoStop}%
\bibitem [{\citenamefont {Keldysh}(1965)}]{Keldysh1965diagram}%
\BibitemOpen
\bibfield  {author} {\bibinfo {author} {\bibfnamefont {L.~V.}\ \bibnamefont
		{Keldysh}},\ }\href@noop {} {\bibfield  {journal} {\bibinfo  {journal} {Sov.
			Phys. JETP}\ }\textbf {\bibinfo {volume} {20}},\ \bibinfo {pages} {1018}
	(\bibinfo {year} {1965})}\BibitemShut {NoStop}%
\end{thebibliography}

%

\end{document}